%% file: Dwarfs_CIV_RB_new_ref_res.tex
\documentclass{emulateapj-rtx4}
\usepackage{apjfonts}
\usepackage{amssymb}
\usepackage{natbib}
 \usepackage{color}
\usepackage{graphicx,mathrsfs,amsmath}
\usepackage[caption=false]{subfig}    % don't load caption or captcont

\shorttitle{The CIV reservoir around dwarf galaxies}
\shortauthors{Bordoloi et al.} 

\newcommand{\lya}{\ensuremath{\rm Ly\alpha\;}}
\newcommand{\rvir}{\ensuremath{\rm R_{vir}}}
\newcommand{\msun}{\ensuremath{\rm M_{\odot}\;}}
\newcommand{\kms}{\rm km~s\ensuremath{^{-1}\;}}
\newcommand{\civ}{\ensuremath{\rm CIV \; 1548 \;}}
\newcommand{\subl}{sub-$\rm{L^*}$}
\newcommand{\lstar}{$L^*$}
\providecommand{\Msun}{\,\ensuremath{\mbox{M}_{\odot}}}
\providecommand{\OVI}{\ensuremath{\mbox{\ion{O}{6$\;$}}}}
\providecommand{\CIV}{\ensuremath{\mbox{\ion{C}{4$\;$}}}}

\newcommand{\hkpc}{h^{-1}{\rm kpc}}
\newcommand{\hmpc}{h^{-1}{\rm Mpc}}

\newcommand{\vw}{{v_{\rm wind}}}
\newcommand{\gad}{{\sc Gadget-2}}

\begin{document}
\title{The COS-Dwarfs Survey: The Carbon Reservoir Around sub-$\rm{L^*}$ Galaxies\altaffilmark{1}}
%\slugcomment{Received ---; Accepted ---}
\author{Rongmon Bordoloi\altaffilmark{2}, 
Jason Tumlinson\altaffilmark{2},
Jessica K. Werk\altaffilmark{3},
Benjamin D. Oppenheimer\altaffilmark{6},  
Molly S. Peeples\altaffilmark{2}, 
J. Xavier Prochaska\altaffilmark{3},
Todd M. Tripp\altaffilmark{4}, 
Neal Katz\altaffilmark{4}, 
Romeel Dav{\'e}\altaffilmark{9,10,11}, 
Andrew J. Fox\altaffilmark{2}, 
Christopher Thom\altaffilmark{2},
Amanda Brady Ford\altaffilmark{5}, 
David H. Weinberg\altaffilmark{7},
Joseph N. Burchett\altaffilmark{4}
\&  Juna A. Kollmeier\altaffilmark{8}
 }
\email{bordoloi@stsci.edu}

\altaffiltext{1}{Based on observations made with the NASA/ESA Hubble
Space Telescope, obtained at
  the Space Telescope Science Institute, which is operated by the
  Association of Universities for Research in Astronomy, Inc., under
  NASA contract NAS 5-26555. These observations are associated with
  program GO12248.}
\altaffiltext{2}{Space Telescope Science Institute, Baltimore, MD}
\altaffiltext{3}{UCO/Lick Observatory, University of California, Santa Cruz, CA} 
\altaffiltext{4}{Department of Astronomy, University of Massachusetts, Amherst, MA} 
\altaffiltext{5}{Steward Observatory, University of Arizona, Tucson, AZ} 
\altaffiltext{6}{Center for Astrophysics and Space Astronomy, Department of Astrophysical and Planetary Sciences, University of Colorado,
389 UCB, Boulder, CO 80309, USA} 
\altaffiltext{7}{Department of Astronomy, The Ohio State University, Columbus, OH}
\altaffiltext{8}{Observatories of the Carnegie Institution of Washington, Pasadena, CA}
\altaffiltext{9}{ University of the Western Cape, Bellville, Cape Town 7535, South Africa}
\altaffiltext{10}{South African Astronomical Observatories, Observatory, Cape Town 7925, South Africa}
\altaffiltext{11}{African Institute for Mathematical Sciences, Muizenberg, Cape Town 7945, South Africa}

\begin{abstract}
We report new observations of circumgalactic gas from the COS-Dwarfs survey, a systematic investigation of the gaseous halos around 43 low-mass $z\; \leq 0.1$ galaxies using background QSOs observed with the Cosmic Origins Spectrograph. From the projected 1D and 2D distribution of \CIV absorption, we find that \CIV is detected out to $\approx$ 100 kpc  (corresponding roughly to $\approx$ 0.5 \rvir) of the host galaxies. The \CIV absorption strength falls off radially as a power law and beyond $\approx$ 0.5 \rvir, no \CIV absorption is detected above our sensitivity limit of $\approx$ 50-100 {m\AA}. We find a tentative correlation between detected \CIV absorption strength and star formation, paralleling the strong correlation seen in highly ionized oxygen for L$\sim${\lstar} galaxies by the COS-Halos survey. The data imply a large carbon reservoir in the CGM of these galaxies, corresponding to a minimum carbon mass of $\gtrsim1.2 \times10^{6}${\Msun} out to $\sim$ 110 kpc. This mass is comparable to the carbon mass in the ISM and exceeds the carbon mass currently in the stars of these galaxies. The \CIV absorption seen around these {\subl} galaxies can account for almost two-thirds of all $W_r \geq $ 100 {m\AA} \CIV absorption detected at low $z$. Comparing the \CIV covering fraction with hydrodynamical simulations, we find that an energy-driven wind model is consistent with the observations whereas a wind model of constant velocity fails to reproduce the CGM  or the galaxy properties. 

\end{abstract}
\keywords{galaxies: evolution,halos--- general---galaxies: quasars: absorption lines--- intergalactic medium}

\section{Introduction}
The distribution and physical conditions of gas around galaxies are important ingredients for understanding how galaxies evolve over cosmic time. Acquiring this knowledge entails studying how the galaxies obtain, process, expel and recycle gas from their surroundings. Understanding these processes is further complicated by the indirect nature of the available observations: stars and gas within the galaxies can be accessed and studied directly, but the gas that is being accreted or being expelled is usually either too diffuse or too hot to be observed directly in emission. Hence we typically have to rely on modeling the outcome (stellar mass, color, morphology) of gas processes rather than examining the phenomena themselves. One of the few ways to advance our understanding of gas processes is to observe the circumgalactic medium (CGM) and to correlate the gas properties with the host galaxies themselves.

Absorption-line systems observed directly in the spectra of background sources provide one of the few effective tracers of the otherwise invisible gas around galaxies. With this technique it is possible to study the distribution and physical conditions of gas around galaxies. The advent of large galaxy surveys \citep{York2000,Lilly2009_Article}, and the installation of the {\it Cosmic Origins Spectrograph} (COS, \citealt{Green2012}), aboard the Hubble Space Telescope (HST) have made it possible to systematically map the CGM and to correlate its properties with the host galaxy properties and the large scale environment \citep{Tumlinson2013,Stocke2013,Bordoloi2011a,Chen2010a,Prochaska2011a, zhu2013a,Werk2013,Stocke14}.

The recent ``COS-Halos'' survey \citep{Tumlinson2013}, mapped the multiphase CGM around \lstar galaxies at $z$ = 0.15 - 0.35. This survey utilized the COS spectrograph to acquire UV spectroscopy of background quasars near 44 $L \gtrsim$ {\lstar} galaxies within a 160 kpc impact parameter \citep{Werk2012}. The COS-Halos survey was optimized to map the multiphase CGM using \OVI and other metal line diagnostics \citep{Tumlinson2011a,Werk2013}. The COS-Halos program finds that \lstar\ star-forming galaxies are surrounded by total CGM gas masses that are comparable to and possibly in excess of their stellar masses  \citep{Werk2014}, and that the metal mass in these halos, is comparable to what has been retained in their interstellar gas and dust \citep{Peeples2013a}. COS-Halos also finds that passive galaxies are relatively deficient in \OVI relative to their star-forming counterparts \citep{Tumlinson2011a}, while the discrepancy in \ion{H}{1} is smaller \citep{Thom2012} and passive galaxies may retain a significant budget of cool CGM gas.

A metals census in the CGM of less luminous galaxies must be carried out at lower redshift to exploit the existing large-scale, flux-limited galaxy redshift surveys such as the Sloan Digital Sky Survey (SDSS, \citealt{Abazajian2009}). However, at $z < 0.1$, the OVI doublet at 1031.93 and 1037.62 {\AA} is not sufficiently redshifted to be efficiently observed with the high resolution COS FUV gratings, which have their best sensitivities at wavelengths greater than 1140 {\AA}. Thus a census of metals at < \lstar must use a tracer that is accessible at $z$ < 0.1 within the COS band. 
 
This paper describes the first results of a new survey (``COS-Dwarfs'') of the CGM gas surrounding a sample of $\rm{L \sim 0.006 - 0.18}$ {\lstar} galaxies in the low-redshift Universe using COS.  This is a direct follow up of the COS-Halos survey in that we focus on the CGM around low-mass galaxies.  Our focus on \subl galaxies is primarily motivated by a desire to extend the mass range of the existing surveys to lower masses and thereby cover a total of $\sim 3$ decades in stellar mass (roughly $M_{*} \approx 10^{8-11}$ \Msun), to better understand the role of the CGM in galaxy formation.

In this paper we focus on the \CIV $\rm{\lambda\lambda}$ 1548, 1550 doublet transition, the most accessible tracer of hot and/or highly ionized gas at redshift $z$ < 0.1. Strong \CIV absorption has been used to trace missing metals in the IGM from $z$ = 4 to today \citep{steidel1990,Danforth2008,DOdorico2010,Simcoe2011,Cooksey2010,Cooksey2013}. This transition has been used extensively at higher redshifts because it is a strong resonant doublet transition giving it a distinct, easily identifiable  characteristic. Also, it is observable redward of the \lya forest, where it becomes easier to identify as it redshifts into the optical passbands at $z$ > 1.5.  \CIV absorption line studies have found that the \CIV mass density relative to the critical density ($\rm{\Omega_{CIV}}$) increases smoothly from $z$ = 4 $\rightarrow$ 1.5 \citep{DOdorico2010} and maps well onto the $z < $1 measurements \citep{Cooksey2010}. $\rm{\Omega_{CIV}}$ increases by a factor of 5 over $z$ = 4 $\rightarrow$ 0, and the comoving \CIV line density $dN_{CIV}/dX$ increases 10-fold over $z\sim$6 $\rightarrow$ 0 for $W_{r} \geq$ 0.6 {\AA} (\citealt{Tilton2012}, \citealt{Cooksey2013}, \citealt{Danforth2014}, Burchett et al., in preparation). This may indicate that the CGM around galaxies is being enriched steadily over the last 12 Gyr lifespan of the Universe. 

However, at high redshifts it is hard to study the individual host galaxies associated with the \CIV absorbers. To study the CGM traced by \CIV around galaxies with well defined properties, in a statistical manner, we rely on HST UV spectroscopy of background quasars, which opens the window of the low-z Universe. The pioneering study of \cite{chen2001} compiled a \CIV-galaxy pair sample of 50 galaxies within 200 kpc, using HST UV spectroscopy of UV bright quasars at $z$ $\approx$ 0.4. They detected \CIV absorption within 100 kpc of their galaxies and found no correlation between absorbers and galaxy morphology or surface brightness.  This study was a ``blind survey'', i.e. it entails taking the spectra of a quasar, finding the absorption line systems first and then looking for associated galaxies in follow up spectroscopic surveys. On the other hand, \cite{Borthakur2013} found, using their small number of absorbers of star-bursting galaxies, that \CIV absorption can be seen even at high impact parameters ($\rm{150 \leq R \leq 200}$ kpc). Individual \CIV systems have also been reported, for which either no L $\gtrsim$ 0.04\lstar galaxies were found within 250 kpc \citep{Tripp2006} or only a very faint dwarf galaxy was detected at $\sim$ 200 kpc \citep{Burchett2013}. 

At higher redshifts, \cite{Adelberger2003,Adelberger2005} observed that almost all of their $ N_{CIV} \ge 10^{14}\; \rm{cm^{-2}}$ absorbers are found within $\approx$ 80 kpc of Lyman-break galaxies (LBGs) at $2 \le z \le 3$. Their findings were based on individual strong absorber-LBG pairs, analysis of stacked spectra of close background galaxies and LBG-\CIV cross-correlation and LBG autocorrelation functions, suggesting that \CIV have the same spatial distribution as the galaxies. \cite{Steidel2010} examined the mean \CIV absorption profile around $z \; \approx$ 2.2 galaxies using stacked background galaxy spectra and found that the mean \CIV absorption strength falls off rapidly at projected distances of 50-100 kpc. \cite{Prochaska2013a} studied the CGM around z $\approx$ 2 bright quasars using closely projected foreground-background quasar pairs. They found that the CGM of quasar hosts exhibit significantly stronger \CIV absorption as compared to $z \; \approx$ 2 LBGs, especially at impact parameters greater than 100 kpc.

In this study, akin to the COS-Halos survey, we adopt a ``galaxy selected'' approach \citep{Bordoloi2011a,Tumlinson2013}. We first identify the projected galaxy-quasar pairs, selected based on the galaxy properties, quasar brightness and the projected separation and then look for absorption in the rest frame of the galaxies. In the ``blind'' method, it is expensive to build up a statistically large sample of quasar-galaxy pairings with  an impact parameter of $<$100 kpc. The galaxy selected approach  allows us to probe the inner part of the CGM by design, enabling us to have better statistics at close impact parameters.  Some of the difficulties that are encountered in correctly associating foreground galaxies with individual absorption systems (e.g.  owing to limitations in the depth of the associated galaxy follow-up) are mitigated by designing a galaxy selected survey around a well defined set of foreground galaxies, rather than starting with a quasar spectrum and trying to identify nearby foreground galaxies responsible for the absorption.

The aim of this paper is to probe the CGM around $z <$ 0.1 low-mass galaxies, allowing us to systematically map the CGM traced by \CIV gas out to their virial radii. Owing to their shallow potential wells, more intense radiation fields and lower metallicities, dwarf galaxies are believed to be inefficient at converting their gas content into stars \citep{Schombert2001}. Dwarf galaxies have higher gas fractions (ISM gas to stellar mass ratio) compared to high mass galaxies \citep{Peeples2013a} and because of their shallower potential wells (compared to the \lstar galaxies), this gas in principle can more easily be expelled into the CGM. Thus, such low-mass galaxies might possess a large reservoir of hidden mass in their CGM. In this work, we estimate the total CGM carbon mass, allowing us to extend the CGM metals census begun by COS-Halos \citep{Peeples2013a} to well below \lstar.

This paper is organized as follows: Section 2 describes the design of the COS-Dwarfs survey. Section 3 describes the data collection and analysis methodology. Section 4.1 presents the 1-D and 2-D \CIV radial absorption profiles. Section 4.2 presents the \CIV covering fraction estimates. Section 4.3 describes the dependence of \CIV absorption strength on galaxy properties and Section 4.4 presents the \CIV absorption kinematics. Section 5 presents the total carbon mass estimate in the CGM of these galaxies. Section 6 shows the expected incidence of \CIV absorbers corresponding to {\subl} galaxy halos. Section 7 compares several theoretical feedback prescriptions with observations. In section 8, we compare our results with the results from previous studies. Finally in Section 9 we summarize our conclusions. 

Throughout our analysis we adopted a cosmology specified by $\rm{\Omega_{m}}$ = 0.238, $\rm{\Omega_{\Lambda}}$ = 0.762, $\rm{H_{0}}$ = 73.2 $\rm{kms^{-1}\; Mpc^{-1}}$, $\rm{\Omega_{b}}$ = 0.0416. Distances and galaxy virial radii are given in proper coordinates.

%----------------------------------------
\begin{figure}[tb!]
\centering
    \includegraphics[height=6.cm,width=7.cm]{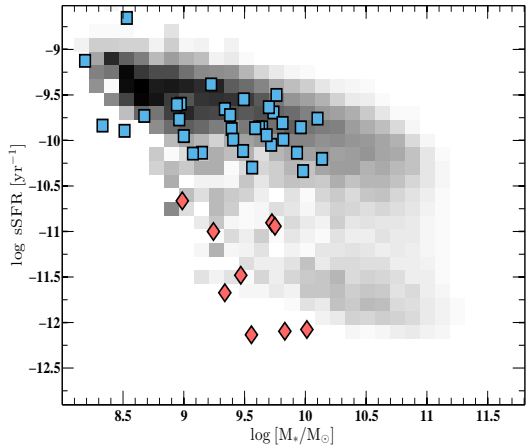}        
    \caption{ The galaxy color-magnitude diagram (sSFR vs $\rm{M_{*}}$) of all COS-Dwarfs galaxies. The blue squares and red diamonds represent star-forming and passive galaxies, respectively. The underlying distribution shows the sSFR vs $\rm{M_{*}}$ for SDSS galaxies from \citep{Schiminovich2007}.}
\label{fig: cmd}
\end{figure}
%----------------------------------------

\section{Survey Motivations and Design} 
Extending the COS-Halos survey to lower stellar mass can reveal how the CGM is configured and evolves as a function of stellar mass, but a new survey was required with design features that differ in many respects from COS-Halos. The requirement to select spectroscopically confirmed galaxies at $\log M_* /M_{\odot} \sim 8-9$ implied that the typical redshift needed to be lower than the median of $ z = 0.25$ in COS-Halos. Another consequence of the redshift range is that COS-Dwarfs galaxies could be selected on the basis of their SDSS {\it spectroscopic} redshifts \citep{Abazajian2009}, rather than photometric redshifts that were employed during the initial selection of the COS-Halos galaxies \citep{Werk2012}. Thus COS-Dwarfs was optimized to yield a sample of galaxies with $\log M_*/M_{\odot} \lesssim 10$  at $z \simeq 0.01 - 0.05$ using Ly$\alpha$, \CIV and a range of other common low and intermediate ion diagnostics. The targeted sample consisted of 39 galaxies with QSO sightlines ranging out to 150 kpc (physical), lying in front of UV-bright QSOs at $z_{QSO} = 0.1 - 1$. In addition to these galaxies, we added four galaxies that were originally selected for COS-Halos but were later omitted on the basis of low luminosity\footnote{These galaxies were all selected for COS-Halos on the basis of photometric redshifts but were found by later spectroscopic measurements \citep{Werk2012} to have redshifts too low, and thus stellar masses too low, to be part of that sample. They are listed in Table 2 of \cite{Tumlinson2013}. }. This yields a total of 43 galaxies that are used for this study. The sample comprises of the closest identified galaxy to each QSO sightline. No other galaxies with known redshifts from SDSS are closer than 300 kpc, though some of the sample galaxies have other galaxies at similar redshifts that are within 300-1000 kpc. We do not consider the sample biased with respect to galaxy neighbors, large-scale environment, or status as a satellite of a larger halo (in cases where neighbors are known). Throughout our analysis we assume that the closest galaxy - i.e. the one selected - is physically associated with the detected absorption. 
 
The galaxy colors and stellar masses $M_*$ were derived from the $ugriz$ SDSS photometry using the template-fitting approach implemented in the \textit{kcorrect} code \citep{Blanton2007} at the measured redshifts. Systematic errors from the mass-to-light ratio and IMF dominate the $\pm$0.2 dex error in stellar mass \citep{Werk2012}. Star formation rates are estimated from the detected nebular emission lines or limited by their absence, with errors up to $\pm$50\%. Fiber losses are corrected using SDSS photometry similar to \cite{Werk2012}. For passive galaxies the SFR is given as a 2$\sigma$ upper limit. Errors on combined quantities, such as the specific star formation rate (sSFR = SFR/$M_*$), are obtained from quadrature sums of the basic error terms. The  specific star formation rates (sSFR) and stellar masses ($\rm{M_{*}}$) of the 43 galaxies used in this study are shown in Figure \ref{fig: cmd}. 
 
The halo masses $M_{halo}$ are computed by interpolating along the abundance matching relation of \cite{Moster2010} at the stellar mass of the galaxy. We scale the projected distance from the sightline to the center of the host galaxy (impact parameter, R) to the virial radius of the galaxy, approximated here as R$_{\rm 200}$, the radius at which the halo mass density is 200 times the critical matter density of the universe. This is given as 
\begin{equation}
			\rm R_{\rm 200}^{    3} = 3 \rm M_{\rm halo}/4\pi\Delta_{\rm vir}\rho_{\rm matter}
\end{equation}
where $\rho_{\rm matter}$ is the critical density at the galaxy redshift times $\Omega_{\rm m}$, and $\Delta_{\rm vir}$ = 200.  At the typical redshifts of the COS-Dwarfs galaxies ($z$ $\sim$ 0.025),  R$_{\rm 200}$ is slightly larger than the virial radius by a factor of $\sim$1.2.  Systematic errors in the stellar mass estimates and the scatter and uncertainty in the M$_{\rm halo}$ - M$_{*}$ relation gives an uncertainty in R$_{\rm 200}$ of approximately 50\%. Throughout this work we refer to this quantity R$_{\rm 200}$ as R$_{\rm vir}$, as is commonly done. We refer the reader to \cite{Werk2012}  for detailed methodology of measuring galaxy properties. All the relevant galaxy properties are tabulated in Table \ref{line_table}.
 
 \section{Data Collection and Analysis}

The data reduction and analysis procedures for the present study follow the procedures described by  \cite{Meiring2011} and \cite{Tumlinson2011b}. In short, we obtain line measurements from 1D spectra that have been merged from individual exposures with different COS UV gratings and central wavelengths. The summation is done in counts per oversampled bin, once the different sub-exposures have been shifted into alignment based on common Milky Way interstellar absorption lines. The photocathode grid wires above the COS microchannel plates, casting shadows onto the detector are the main source of fixed-pattern noise in our data. Smaller fluctuations caused by the microchannel plate pores generally do not appear at the S/N ratios of our data. There are other fixed-pattern noise features that must also be removed. We adopted flat-field reference files prepared and communicated to us by D. Massa at STScI and filtered for high-frequency noise by E. Jenkins. These 1D files allow us to correct the shadowed pixels by modifying the effective exposure time and count rate in each pixel prior to coadding it with the others.  These flats do not correct for depressions in COS spectra caused by gain sag in some regions of the detector after prolonged exposure to bright geocoronal emission lines \citep{Sahnow2011}. These gain-sag features can mimic real absorption lines, but the affected regions are flagged by the CALCOS pipeline \citep{COSHandbook2012}.  To avoid creating bogus absorption features, we reject the flagged gain-sag regions in the coaddition process. The resulting 1D, flat-corrected spectra are binned to Nyquist sampling with two bins per resolution element and a S/N of $\sim$ 10-12 per COS resolution element (FWHM $\simeq$ 18 {\kms}). The fully reduced spectra are in units of counts per second with appropriate errors from counting statistics (Poisson) and errors propagated from the CALCOS \citep{COSHandbook2012} calibration steps. 

With fully reduced spectra, we proceed with a semi-automated framework developed for COS-Halos to identify and measure the  absorption lines associated with the foreground galaxies. We refer the reader to \cite{Tumlinson2013} for a detailed description of the method and only briefly summarize the steps here. We shift the reduced, co-added quasar COS spectrum to the rest-frame of the foreground galaxy, setting $v=0$ \kms to the systemic redshift of the foreground galaxy. The systemic redshifts of the galaxies are measured to high precision ($\sigma_{specz} \sim 30$ {\kms} in the rest-frame) and we can focus on common lines at predictable places in observed wavelength. A detected absorption line is considered to be associated with a galaxy if it is within $\pm$ 600 \kms of the systemic redshift of the galaxy. Short slices of data within $\pm$ 600 \kms of the systemic redshift of the galaxy are extracted around each line of interest, including the \CIV doublet and $\rm{ Ly\alpha}$, CII 1334, SiIII 1206, SiIV 1393, SiIV 1402. Each slice is independently continuum-normalized using a fifth-order Legendre polynomial, and the equivalent width and column density of the absorption is measured (as a detection or an upper limit for non-detections) using the apparent optical depth (AOD) method of \cite{Savage1991}. For the key ion C IV, we visually inspect each doublet to confirm their presence and search for contamination, and to set the velocity range for AOD integration. To minimize contamination from other intervening absorption line systems or from the foreground ISM of the Milky Way, we attempt to identify every detection feature within a $\pm$ 600 \kms range at the position of the \CIV absorption doublet. Most such features are not associated with the target galaxy and are positively identified HI or metal lines associated with other absorbers at different redshifts along the line of sight. We confirm the presence of the \CIV absorption doublet by excluding any alternate identification for the doublet and by looking at their apparent optical depth profiles. We accept apparent optical depth ratios ranging from 2:1 to 1:1 and require that the \CIV absorption is aligned within $\pm$ 200 \kms of \lya and other detected metal lines in that absorption system. 

In addition to the AOD-derived column densities, we fit Voigt profiles to both absorption lines of the \CIV doublet, to assess kinematic component structure and to estimate column densities for severely blended lines. The profile fitting improves on direct line integration of column densities using information about line shape and placement to constrain the fit. It also helps assess line saturation by taking into account the true line-spread function (LSF) of the instrument.  We refer the reader to \cite{Tumlinson2013} for a detailed description of the Voigt profile fitting method. To summarize, we use an IDL iterative fitting program using the MPFIT software to optimize the fit and to estimate the errors on the fits. We fit column density $N$, Doppler $ b$ parameter, and velocity offset $v$ for each component simultaneously for each \CIV doublet. The initial parameters, including the number of components to be fitted, are set after visual examination of the data.  The best fit parameters are obtained by performing $\chi^{2}$ minimization around $\pm$ 600 \kms of the foreground galaxy systemic redshift using the Levenberg-Marquardt algorithm. The detected \CIV absorption lines and the corresponding Voigt profile fits are shown in the Appendix (Figure \ref{fig:cont}). \CIV absorption at these redshifts are detected on the G160M grating of the COS spectrograph. For each individual quasar spectrum, we compute the S/N and 3$\sigma$ detection threshold at the \CIV rest frame. The quoted 3$\sigma$ detection thresholds are the mean uncertainty in measuring the \CIV rest frame equivalent width with a 100 \kms velocity window, within $\pm$600 \kms of the \CIV line. All the measurements and galaxy properties are summarized in Table \ref{line_table}.

%----------
\begin{figure*}[htb!]
\centering
    \includegraphics[height=6.cm,width=7.cm]{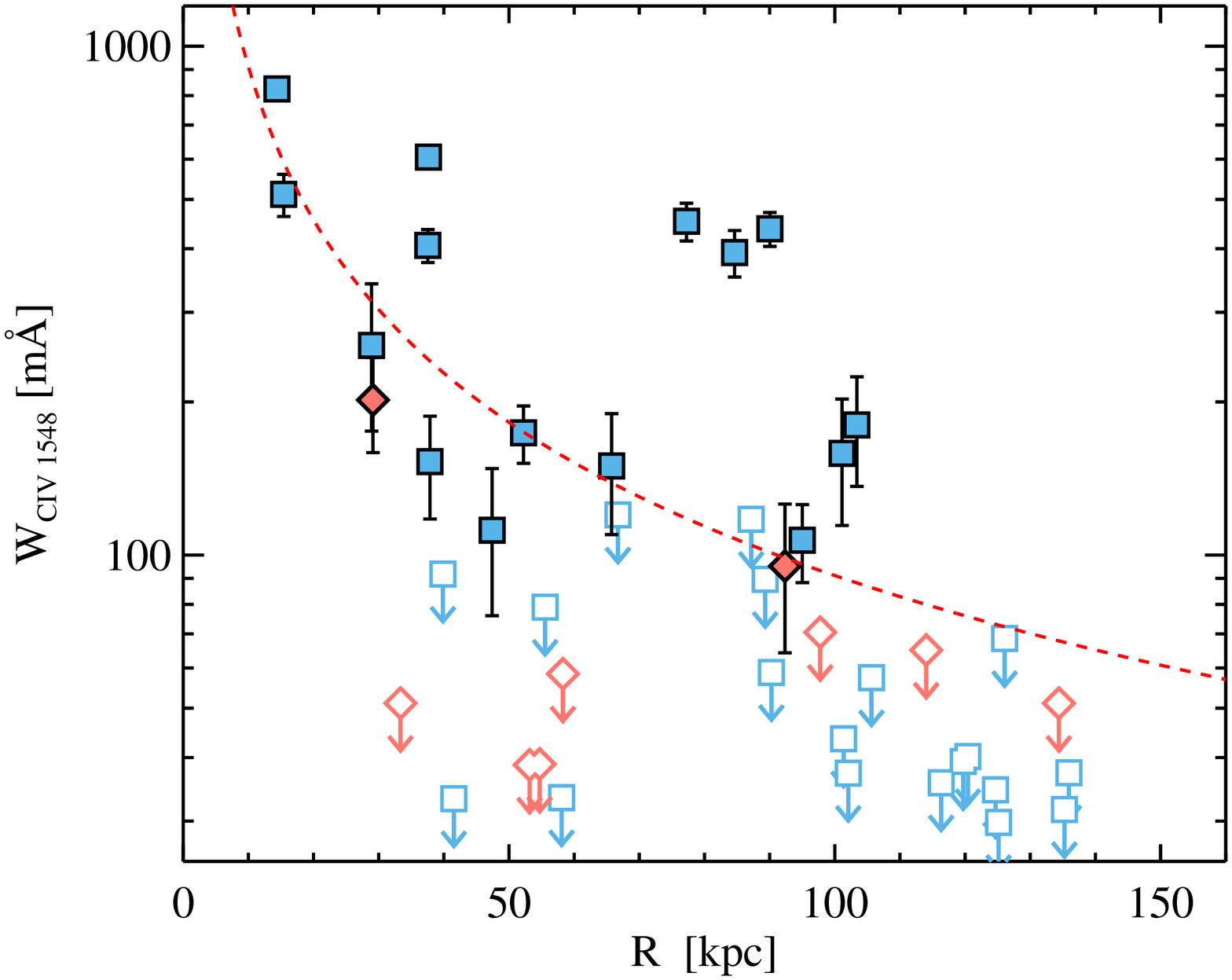} 
     \includegraphics[height=6.cm,width=7.cm]{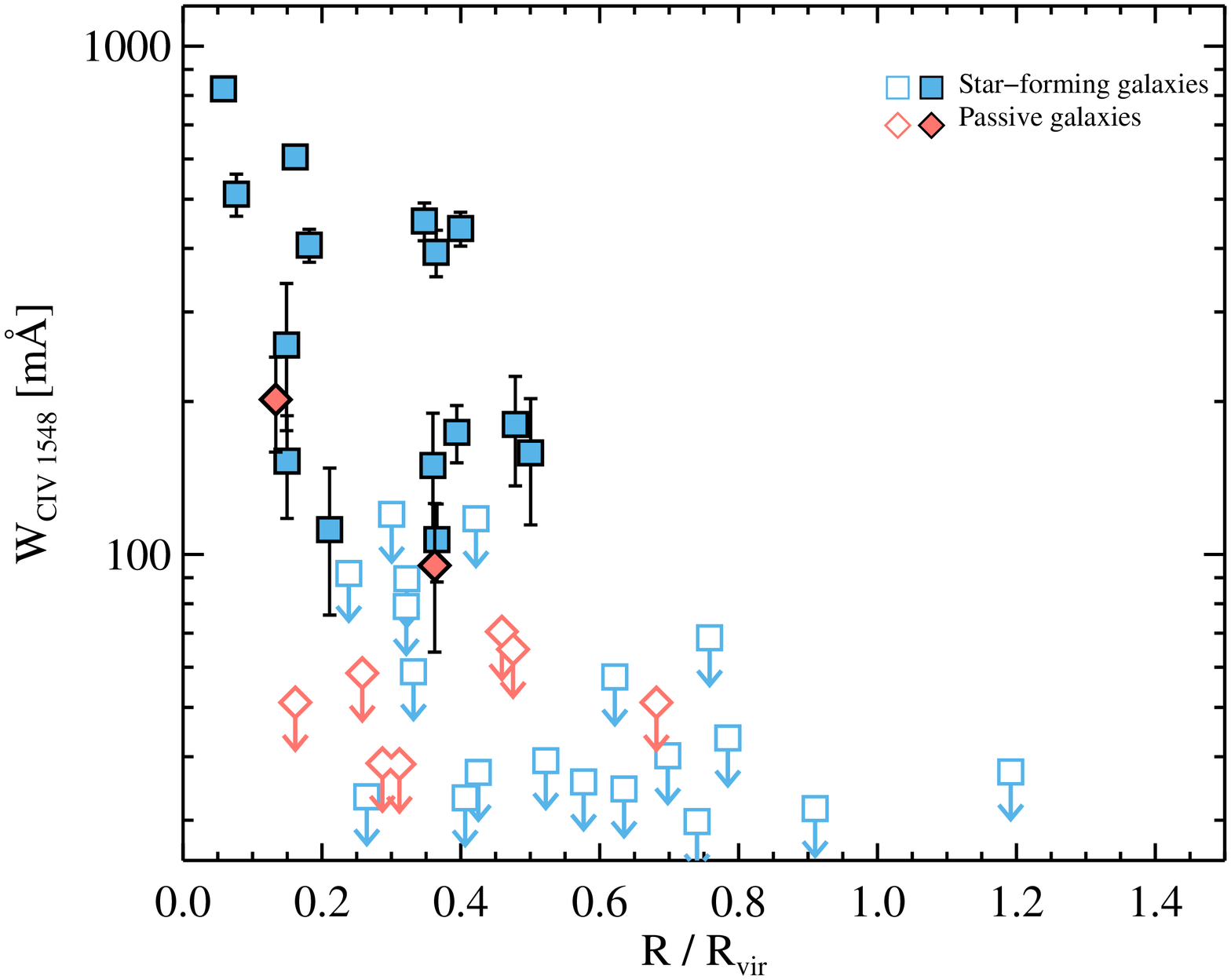}
      \includegraphics[height=6.cm,width=7.cm]{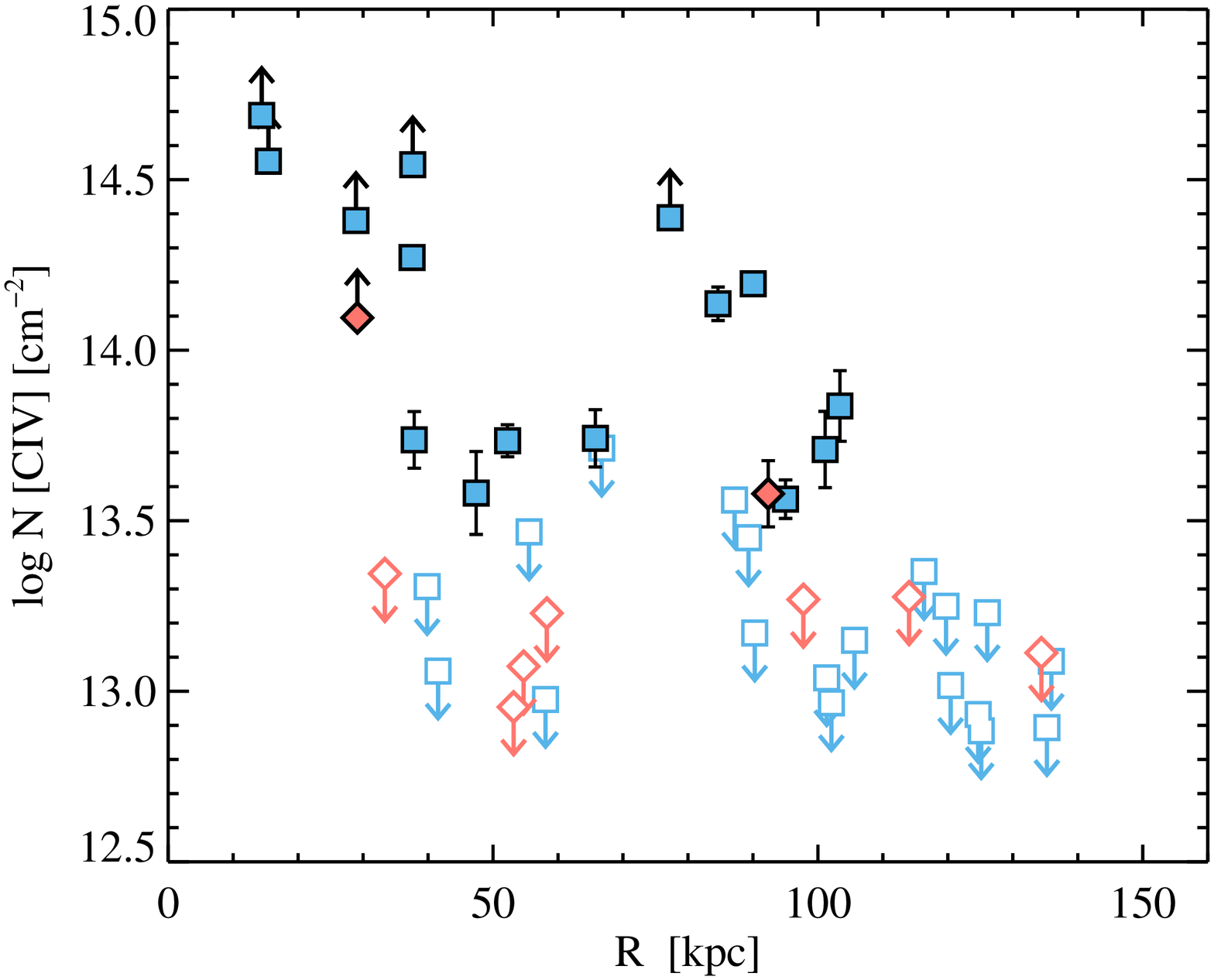}   
     \includegraphics[height=6.cm,width=7cm]{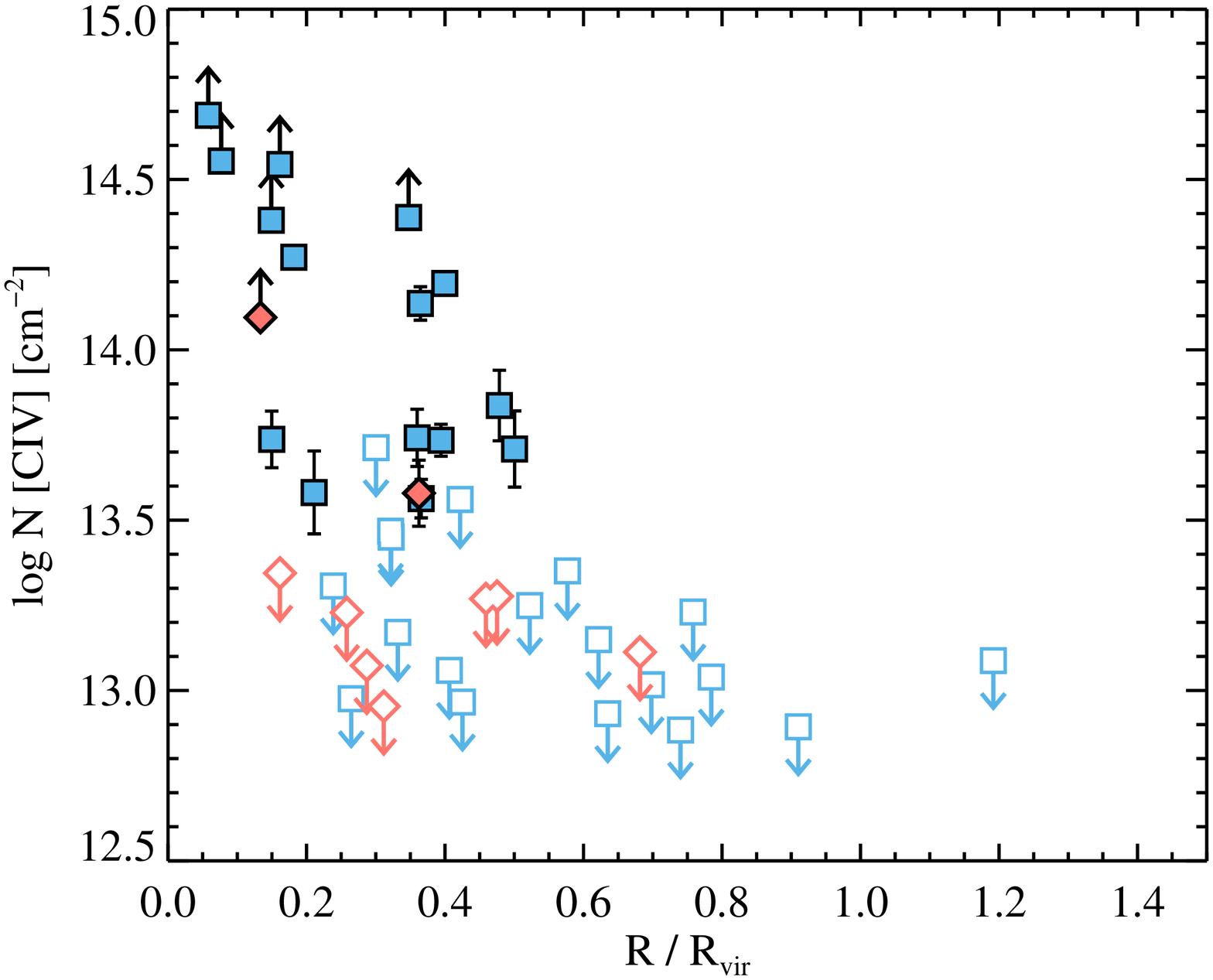}      
\caption{The 1-D \CIV absorption profile around the COS-Dwarfs galaxies, in terms of R (left panels) and {\rvir} (right panels). The blue squares and red diamonds represent star-forming and passive galaxies respectively. The filled symbols indicate detections while the open points with arrows indicate 2$\sigma$ upper limits of non-detections. Absorption strength declines with increasing projected galactocentric radius and no \CIV absorption is detected beyond R $\gtrsim$ 110 kpc. The dashed red line represents the best fit radial profile to the data given by equation \ref{radial_profile}. Bottom Panels show the \CIV absorption radial profile in terms of AOD column densities. The filled symbols with upward arrows are lower limits on \CIV column densities.} 
\label{fig: 1d radial profile}
\end{figure*}

\begin{figure*}[htb!]
\centering
    \includegraphics[height=6.cm,width=7.cm]{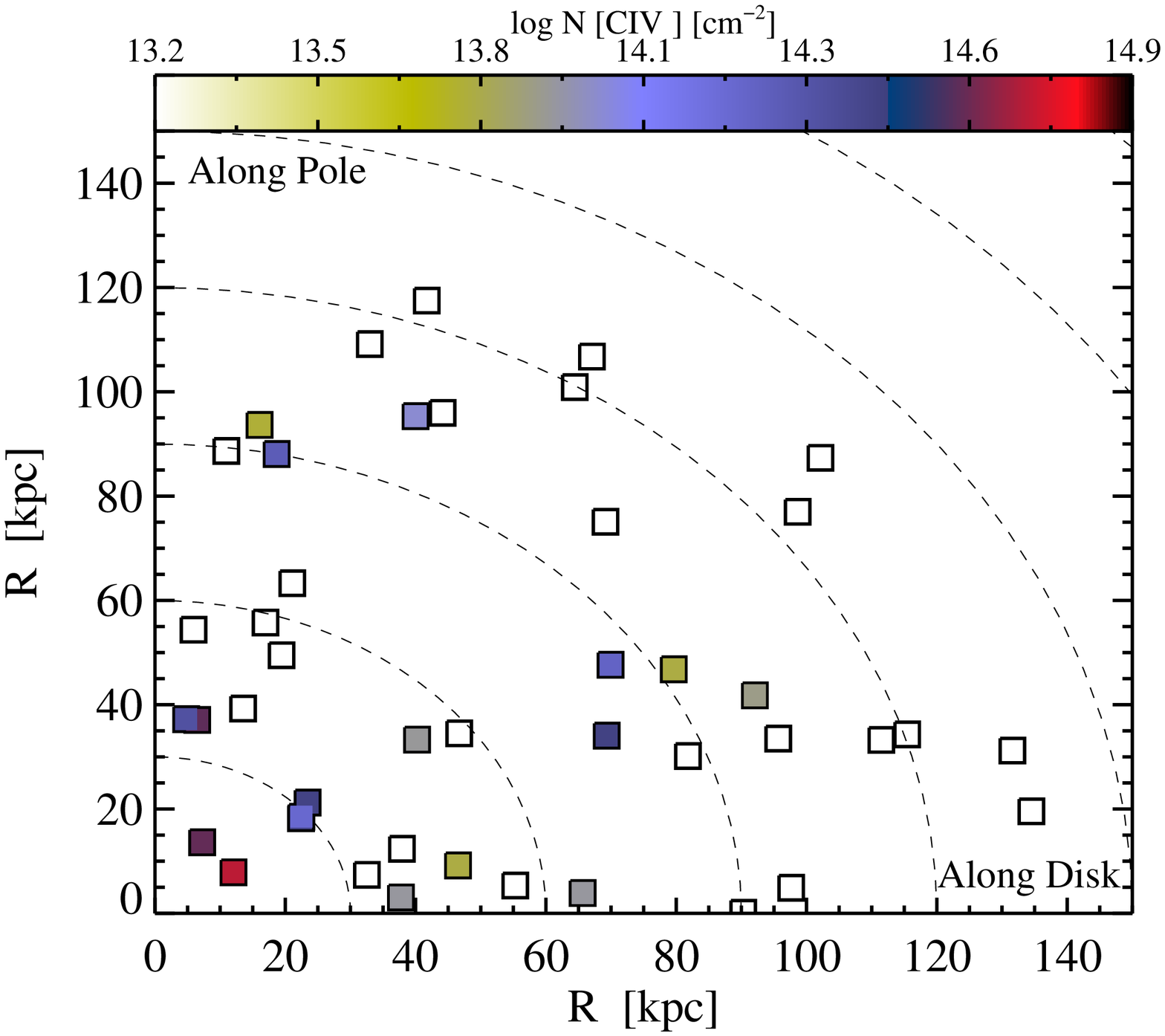}       
     \includegraphics[height=6.cm,width=7.cm]{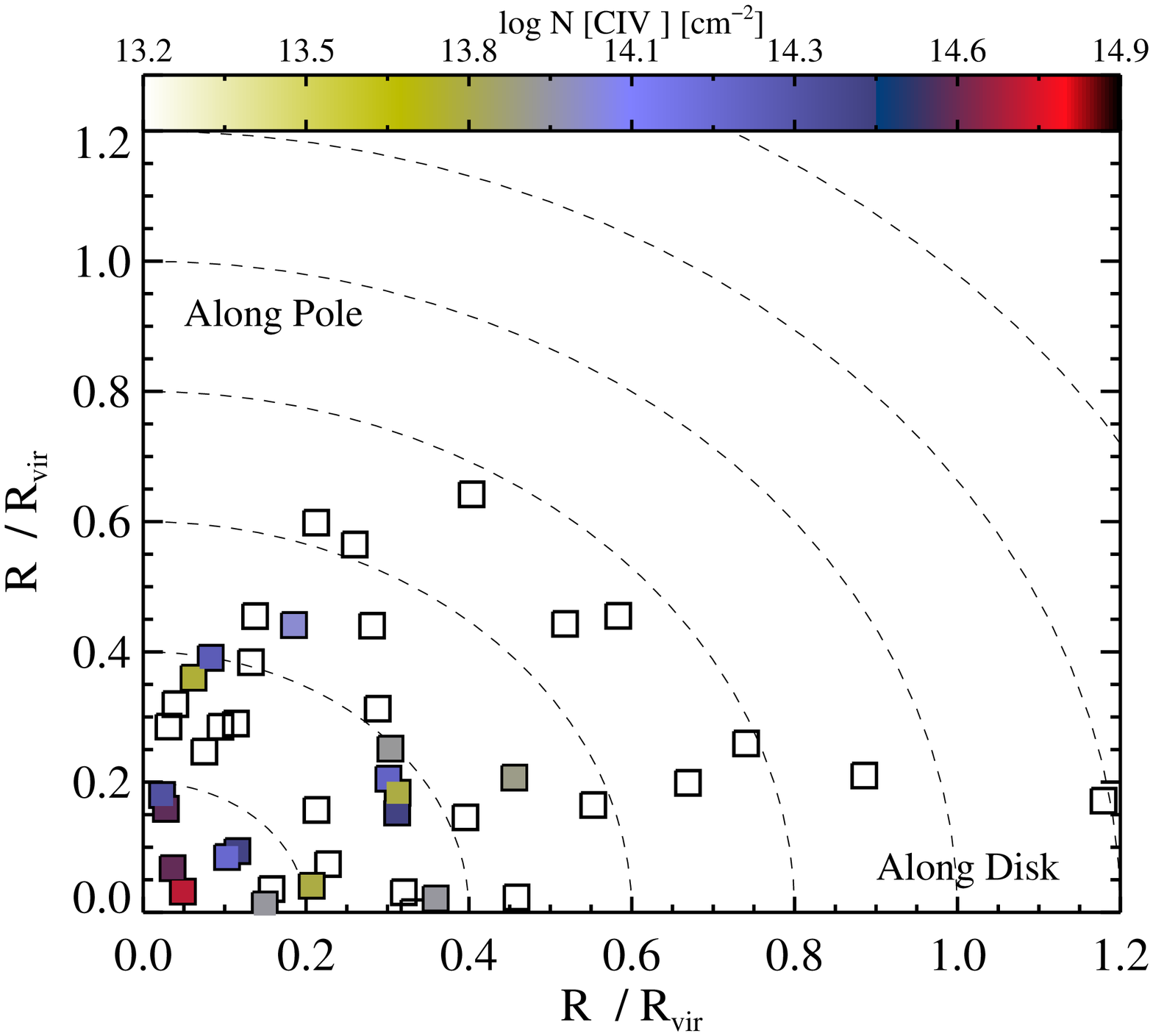}            
    \caption{The projected  2-D \CIV absorption radial profile around the COS-Dwarfs galaxies, in terms of R (left panel) and {\rvir} (right panel). The open squares indicate non-detections and the filled squares are  detections. The detections are color coded to reflect their AOD column densities.  The points lying along the y-axis represent lines of sight passing along the projected minor axis of the galaxy and the points lying along the x-axis represent lines of sight passing along the projected major axis of the galaxy.}
\label{fig: 2d radial profile}
\end{figure*}

%----------

\begin{figure*}[htb!]
\centering
    \includegraphics[height=6.cm,width=7.cm]{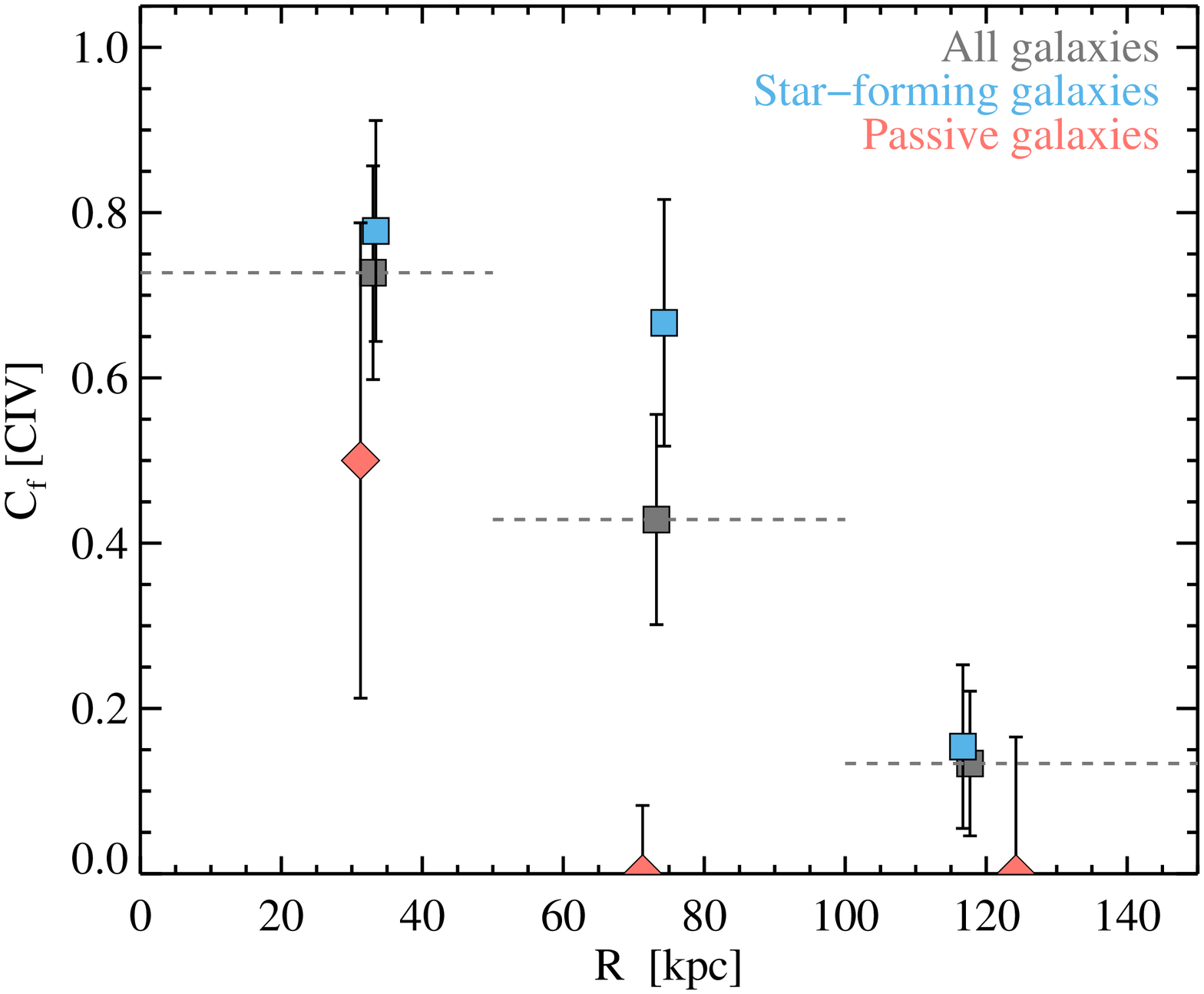}      
        \includegraphics[height=6.cm,width=7.cm]{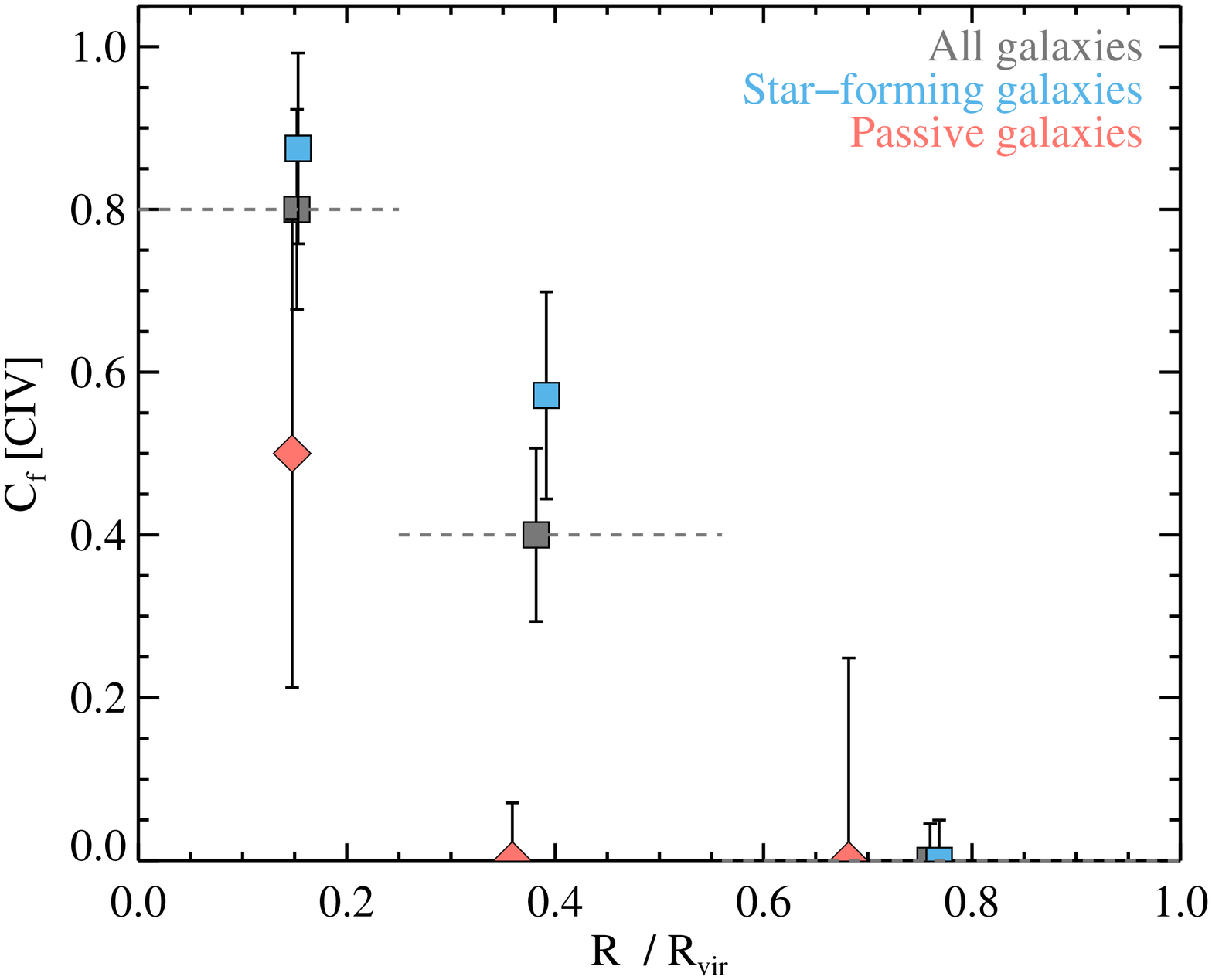}     
         \includegraphics[height=6.cm,width=7.cm]{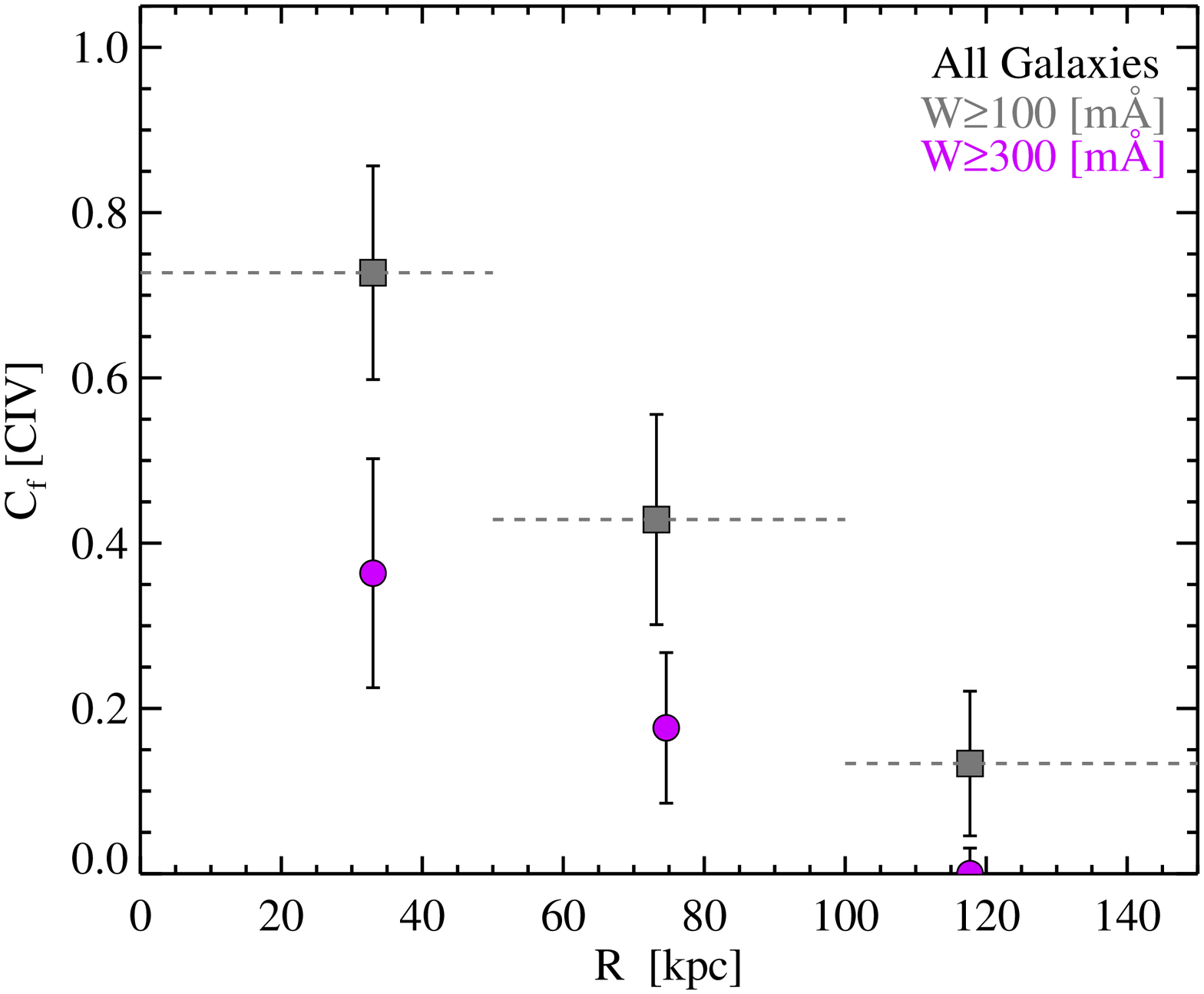} 
          \includegraphics[height=6.cm,width=7.cm]{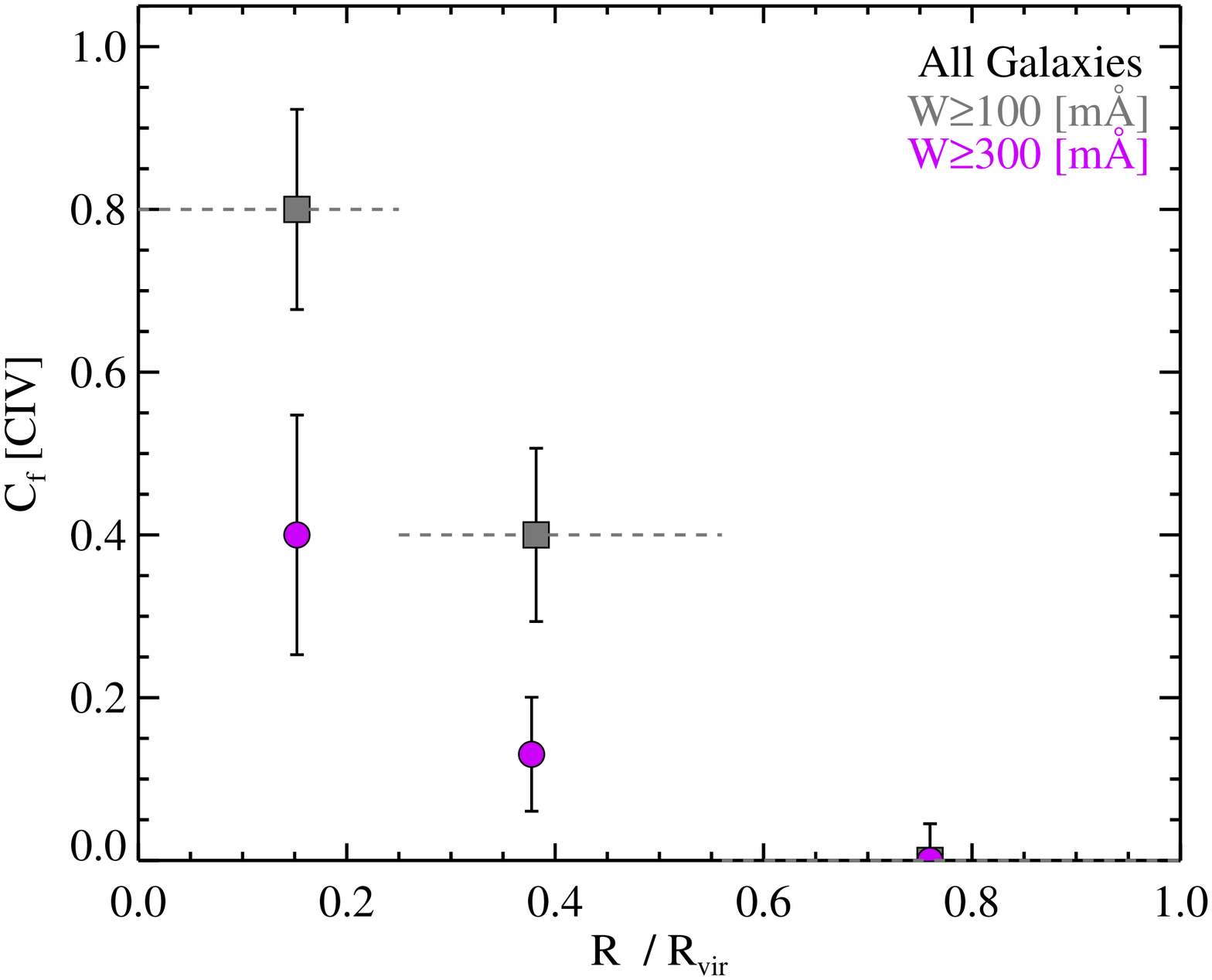}       
     
    \caption{The \CIV absorption covering fraction around galaxies, in terms of R (left panels) and {\rvir} (right panels). The error bars on covering fraction represent the 68\% confidence intervals. The horizontal dashed lines are range bars indicating the width of the radial bins. Top panels: The blue squares and red diamonds are covering fraction estimates for star-forming and passive galaxies, respectively, while the gray squares are covering fractions for all galaxies with $W \geq $100 m{\AA}.  The covering fraction drops sharply beyond {0.2\rvir} and is zero at \rvir $\gtrsim$ 0.5. Even very close to the galaxies, the \CIV covering fraction is not unity.  Bottom Panels: Covering fraction for \CIV absorption around galaxies for $W \geq 100$ m{\AA} (gray squares) and $W \geq 300$ m{\AA} (blue circles),  respectively. }
\label{fig: covering fraction}
\end{figure*}

\section{Results}
In the following sections we discuss the variation of \CIV absorption with different properties of the foreground galaxies, examining the radial profile, the dependence on galaxy sSFR, and absorber kinematics.

\subsection{The Spatial Distribution of \CIV Absorption Around Galaxies}
  
To estimate the spatial extent of \CIV absorption around these galaxies, we examine how the \CIV rest-frame equivalent width ($W_r$) and column density ($\rm{N_{CIV}}$) depend on the impact parameters (R) and the virial radii {\rvir} of the host galaxies. Figure \ref{fig: 1d radial profile} shows the 1-D \CIV absorption profile around 43 galaxies as a function of R (left panels) and {\rvir} (right panels). \CIV absorption is seen around 17 galaxies (filled symbols), and no absorption is detected around 26 galaxies (open symbols). A galaxy is defined to be star-forming (blue squares) if its specific star formation rate (sSFR) is greater than $\rm{\log sSFR \geq -10.6}$.  

Figure \ref{fig: 1d radial profile} reveals that the strongest \CIV absorption is detected at smaller impact parameters. The CGM around these galaxies shows a drop in \CIV absorption beyond a radius of $\sim$ 110 kpc ($\gtrsim$0.5 \rvir), (see also \citealt{JiaLiang2014} for a similar finding).  All $\log  N_{CIV} \ge 14$ absorption lines are detected within $\approx$ 90 kpc of the galaxies. We parametrize the radial fall off in \CIV absorption by characterizing the 1D radial profile in Figure \ref{fig: 1d radial profile}, in terms of a power law with an exponential fall off. This characterization is given as 
			
\begin{equation}
\rm{ \langle W_{r}(R) \rangle  = W_{0}\times \Big( \frac{R}{R_{x}} \Big)^{-1} \times \exp \Big( \frac{-R_{vir}}{R_{x}}  \Big)
}
\label{radial_profile}
\end{equation} 	
Here the characteristic radius $\rm{R_{x}}$ sets the power of the exponential fall off for a galaxy with a given \rvir. We perform a maximum likelihood fit to the data to obtain the best fit parameters as $\rm{R_{x} = 105 \pm 27}$ kpc and, $\rm{W_{0} = 0.6 \pm 0.3 }${\AA}. The fitted mean absorption profile with $\rm{\langle R_{vir} \rangle }$ =203 kpc is the dashed red line in the top left panel of Figure \ref{fig: 1d radial profile}. 

Figure \ref{fig: 1d radial profile} further reveals that the {\CIV} absorption is patchy even at low impact parameters. At R $ \leq$ 60 kpc, 53\% (9 out of 17) of galaxies are associated with \CIV absorption while 40\% of the entire sample (17 out of 43) is associated with {\CIV} absorption ($W_r \;\geq$ 90 {m\AA}). Even after excluding three non-detections with > 90 {m\AA} 3$\sigma$ detection thresholds (Table \ref{line_table}), we find that 43\% of the entire sample (17 out of 40) is associated with {\CIV} absorption. The patchiness is also evident in the scatter of {\CIV} absorption strength for the detected absorbers (standard deviation = 208 {m\AA}).

For star-forming galaxies, the COS-Dwarfs data clearly shows a trend between virial radius (\rvir) and \CIV absorption strength, that is quite similar to that observed for Mg II absorption (e.g. \citealt{Chen2010a}, \citealt{Bordoloi2011a}, \citealt{Werk2013}). As we probe the innermost CGM at small virial radii, increasingly stronger \CIV absorbers are detected. This is different than the radial profile observed for \OVI \citep{Tumlinson2011a}, where the {\OVI} radial profile for star-forming galaxies is flat out to 150 kpc and there are very few non-detections. The observed \CIV absorption profile is also not consistent with the \OVI absorption profile observed around $>$0.1\lstar galaxies at $z <$0.2 \citep{Prochaska2011c}. They detected extended ($\approx$200 kpc) \OVI absorption around  $>$0.1\lstar galaxies with high covering fraction ($\approx$ 80\%). However, \cite{Prochaska2011c} found that, their small sample of dwarf galaxies ($<$0.1\lstar), exhibit much lower \OVI covering fraction for R $>$ 50 kpc at $\langle z \rangle \; \approx$ 0.04. It should be noted that \OVI and \CIV have different ionization potentials, and hence they trace CGM gas with different physical conditions. In particular, \OVI is believed to trace more diffuse gas in the CGM compared to \CIV. Part of the discrepancy between the \CIV and \OVI radial profiles could owe to different species tracing gases with different physical conditions.  However, some absorbers show a remarkable kinematical correlation of the velocity centroids and line widths of species ranging from H I, Mg II, and C II up to \OVI and NeVIII \citep{Tripp2008, Tripp2011, Meiring2013,Savage2014}.  In these absorbers, models indicate that the low and high-ionization species must arise in different phases, but somehow the different phases are forced to move together and likely originate in the same general spatial region.  Unfortunately, most current low-z absorber samples do not provide simultaneous coverage of \CIV and \OVI, but in a small number of cases where \CIV and \OVI are both observed, they are often found to be well-aligned and have essentially identical kinematics \citep{Tripp2006, Savage2014}. Thus, the different radial profiles from COS-Dwarfs and COS-Halos could alternatively be an intrinsic difference between \subl and \lstar galaxies.  

\subsubsection{Two-dimensional \CIV absorption profile}
Since these low redshift galaxies have well resolved light profiles from ground based imaging, we use SDSS r band photometric measurements for each foreground galaxy to study the azimuthal dependence of \CIV absorption. We compute the azimuthal angle ($\phi$), the projected angle that the quasar line of sight makes with the projected major axis of the galaxy, by using SExtractor \citep{Bertin1996} to estimate the position angle and the axis ratio. The azimuthal angles are presented in Table \ref{line_table}. The sense of the azimuthal angle is such that any sightline passing along the projected major axis of the galaxy is assigned $\phi$ = 90$^{\circ}$ and any sight line passing along the projected minor axis of the galaxy is assigned $\phi$ = 0$^{\circ}$. 

Figure \ref{fig: 2d radial profile} shows the projected 2-D \CIV absorption profile around the COS-Dwarfs galaxies in terms of R and {\rvir} respectively. The open squares are the non-detections and the filled squares are color coded to reflect their absorption strengths. Visual inspection of Figure \ref{fig: 2d radial profile} shows that stronger absorbers have a marginal preference to be aligned towards the minor axis of the galaxies, however this trend is not statistically significant. A two sample KS test does not rule out the null hypothesis that the \CIV absorbers are spherically distributed around the galaxies at 10\% significance. Comparing the azimuthal asymmetry of \CIV and Mg II absorption line systems \citep{Bordoloi2011a,Bordoloi2012a} will allow us to constrain the geometry of the multi phase CGM in a unique way. The strong Mg II absorbers are preferentially seen along the projected minor axis of disk galaxies. However, if the observed \CIV detections represent any azimuthal asymmetry, we would need $\approx$ 110 absorber-galaxy pairs to rule out the null hypothesis that \CIV absorption is uniformly distributed around the galaxies at 1\% significance.

\subsection{\CIV Covering Fraction Estimates} 
In the previous subsection, we have shown how the \CIV absorption strength changes as we probe the CGM away from the galaxy. We have also shown that the number of non-detections increases as we probe higher projected distances. Here we quantify the observed incidence of \CIV absorption around galaxies in terms of covering fraction ($C_{f}$). The covering fraction is defined as 

\begin{equation}
C_{f} = \frac{N_{W \geq W_{cut}}(R)}{N_{tot}(R)} ,
\end{equation}
where, $N_{W \geq W_{cut}}(R)$ is the number of lines of sight within a given radial bin from the galaxies, having associated absorbers of strength $W$ greater than some cutoff equivalent width ($W_{cut}$). $N_{tot}(R)$ is the total number of lines of sight in that same radial bin. 

The typical detection threshold for the COS-Dwarfs survey is $\approx$ 50-100 {m\AA}. Throughout the paper, we will compute covering fraction only for absorption with $W_{cut} \geq$ 100 m{\AA}.  If we exclude the three non-detection with 3$\sigma$ detection threshold >100 m{\AA} from the sample (Table \ref{line_table}), $C_{f}$ for the complete sample is 40\% (16 detections out of 40).

We first evaluate $C_{f}$ for all star-forming and passive galaxies including all absorbers with $W_{cut} \geq$ 100 m{\AA}. In Figure \ref{fig: covering fraction} we show the $C_{f}$ estimates as a function of R (top left panel) and $\rvir$ (top right panel). Star-forming galaxies (blue squares) exhibit a statistically higher covering fraction than passive galaxies (red diamonds). The gray squares represent the $C_{f}$ estimates for all galaxies. The error bars are 68\% confidence intervals on the $C_{f}$ estimates. At close projected galactocentric distances the star-forming galaxies have very high covering fractions ($\approx$ 80\%) that decrease to $\sim$ 65\% at R $ \sim$ 80 kpc or \rvir $\sim$ 0.4. For R $ > $100 kpc the covering fractions are very low (15\%) and, in terms of \rvir, $C_{f}$ goes to zero as we probe beyond \rvir $\gtrsim$ 0.5. The horizontal dashed lines are range bars indicating the span of the radial bins used to compute the covering fractions.

We further explore how $C_{f}$ varies for a different values of $W_{cut}$ in the bottom panels of Figure \ref{fig: covering fraction}. The  $C_{f}$ estimates as a function of R (bottom left panel) and $\rvir$ (bottom right panel) are shown for $W_{cut} \geq 300$ m{\AA} (purple circles) and  $W_{cut} \geq 100$ m{\AA} (gray diamonds), respectively. At close galactocentric radius, the $C_{f}$ for ``strong'' \CIV absorption ($W_{cut} \geq 300$ m{\AA}) is about half ($\sim$ 40 \%) of the $C_{f}$ for absorbers with $W_{cut} \geq 100$ m{\AA}. In both cases, the radial fall off in $C_{f}$ with distance exhibits a similar trend. 

The \CIV absorption is seen out to $\sim$ 110 kpc in these galaxies ($\approx$ 0.5 {\rvir} in terms of virial radius), beyond which \CIV absorption is not detected in any of the sightlines (0 out of 11). This suggests that the \CIV covering fraction falls off sharply at 0.5{\rvir}, beyond which there is little strong \CIV absorption (see also \citealt{JiaLiang2014}). This sharp cutoff is quite similar to that seen in MgII absorption line systems \citep{Chen2010a,Bordoloi2011a,Nielsen2012}.

\begin{figure*}[t!]
\centering
    \includegraphics[height=6.cm,width=7cm]{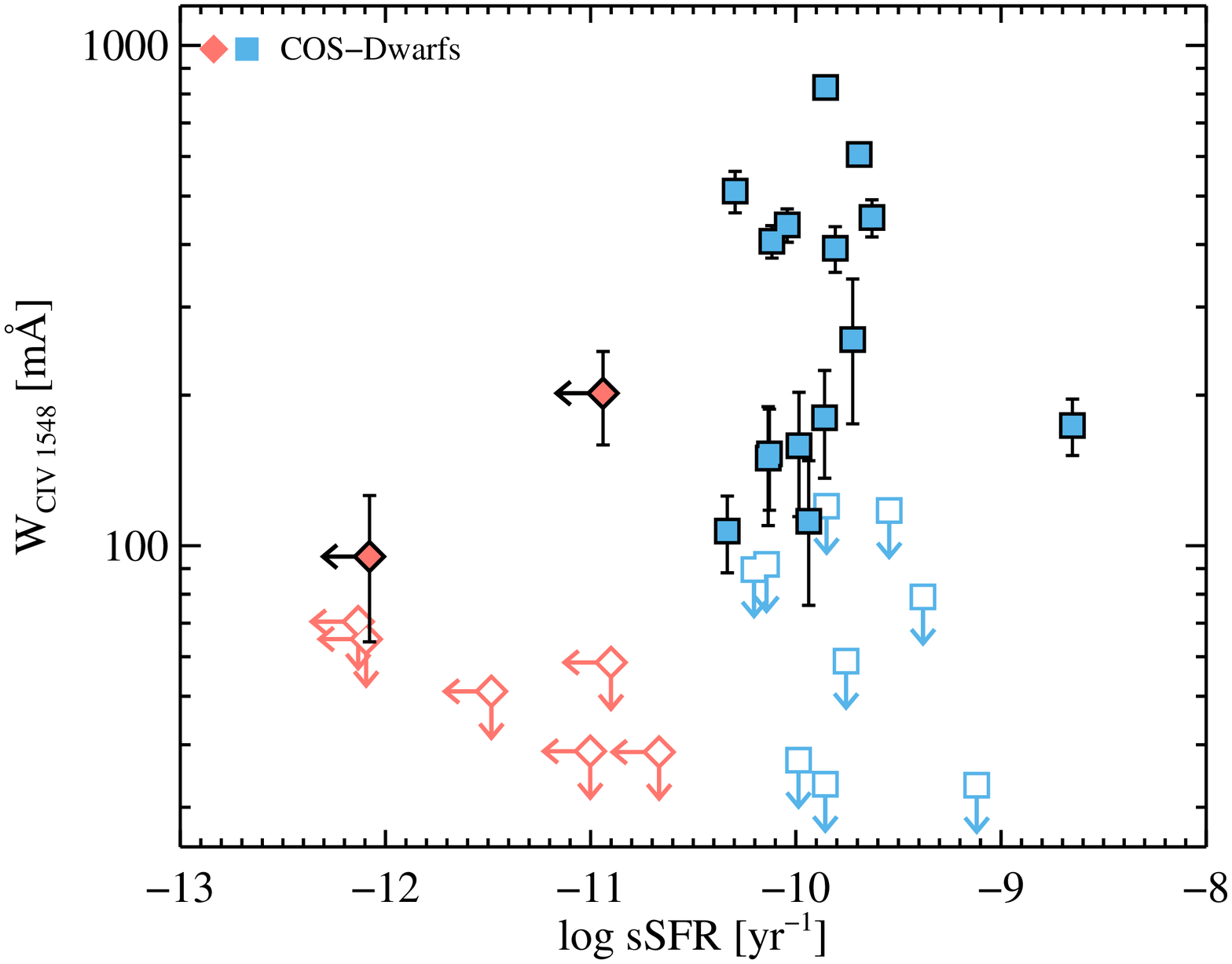}    
    \includegraphics[height=6.cm,width=7cm]{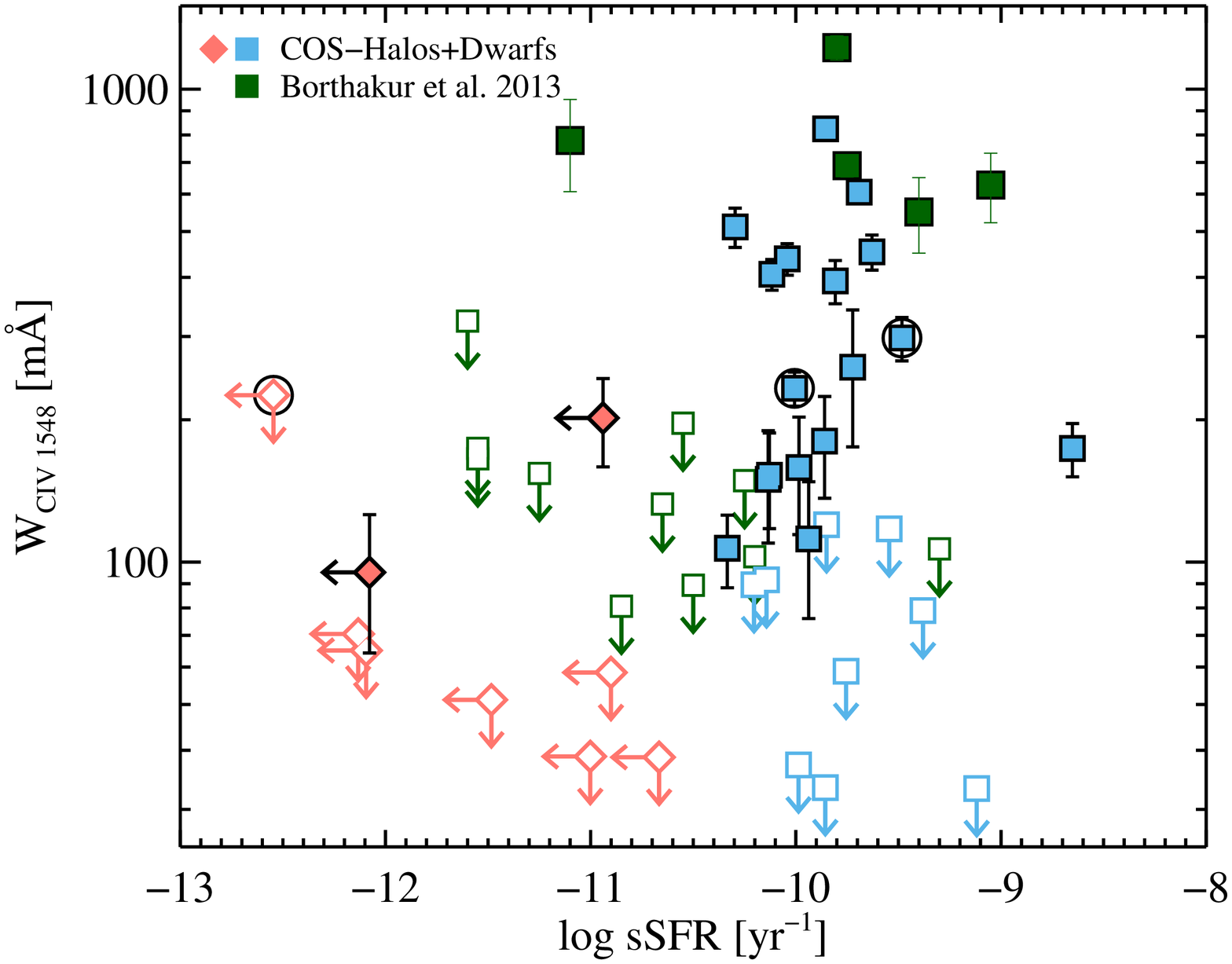}
    \includegraphics[height=6.cm,width=7cm]{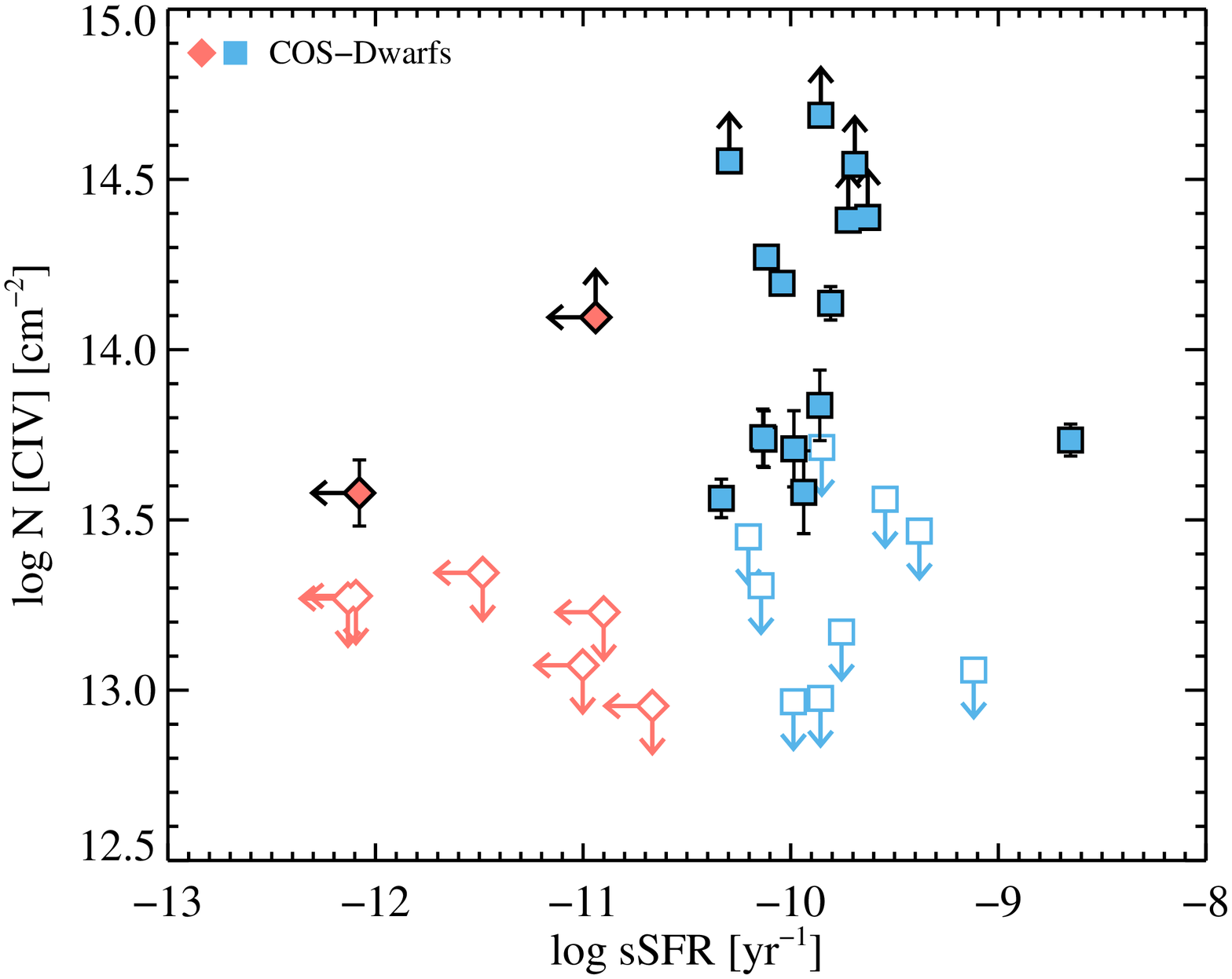}    
    \includegraphics[height=6.cm,width=7cm]{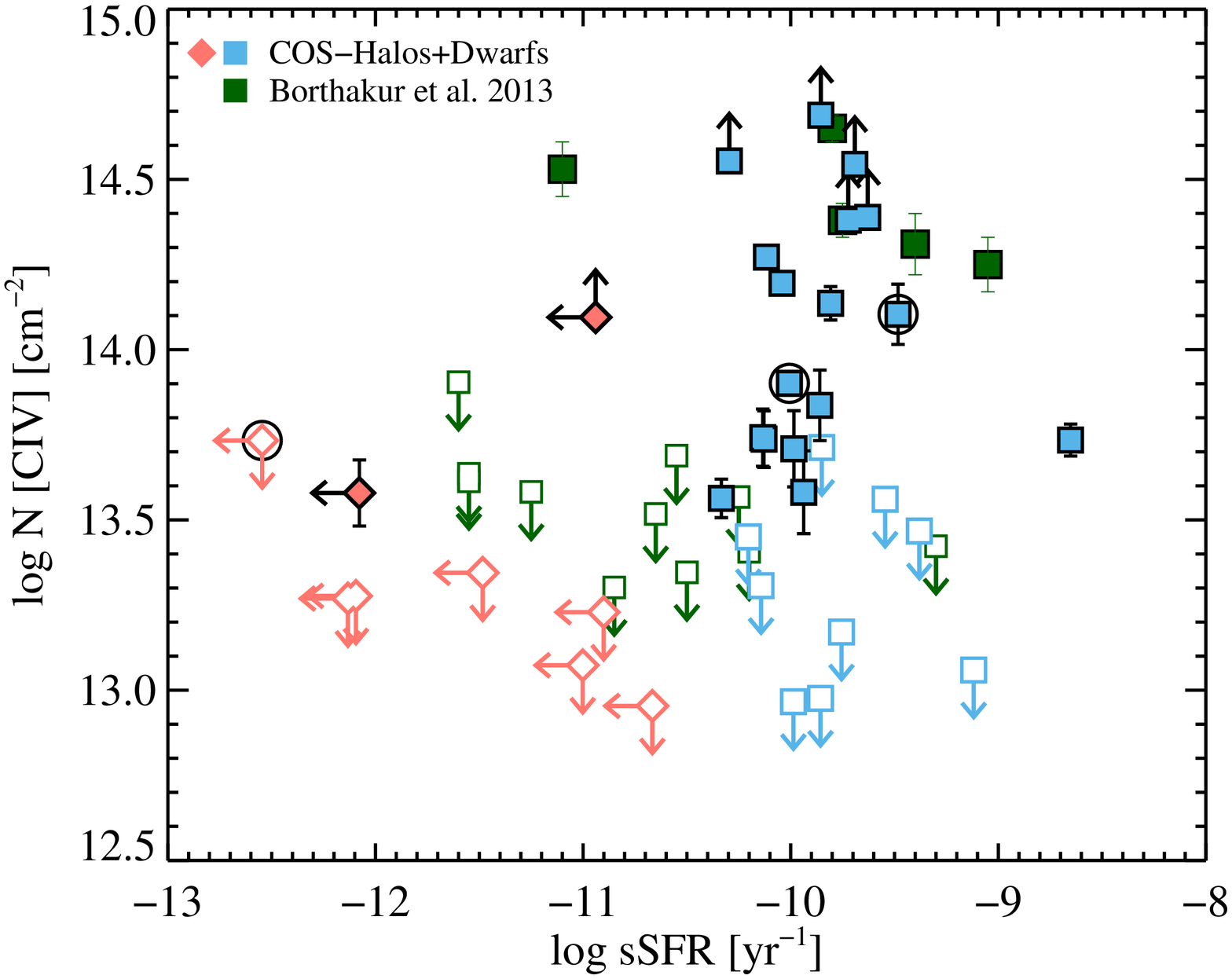}    
    \caption{Top Panels: Dependence of \CIV rest frame equivalent width on the sSFR of the host galaxies within \rvir $\leq$ 0.5 from the COS-Dwarfs survey (left panel). The right panel shows the same after including the data from COS-Halos (tagged with open circles) and, \cite{Borthakur2013} (green points). The star-forming (blue squares) and passive (red diamonds) galaxies show distinctly different \CIV detection rates. The filled symbols represent detected \CIV absorption and the open symbols with arrows are 2$\sigma$ upper limits on non-detections.  Bottom Panels: Same as the top panels, now in terms of AOD column densities. The open symbols with downward arrows are 2$\sigma$ upper limits on non-detections and the filled symbols with upward arrows are lower limits on column densities.}
\label{fig: color ssfr}
\end{figure*}

\subsection{Dependence of \CIV absorption with galaxy sSFR}
To characterize how \CIV absorption depends on host galaxy properties, we study the variation of \CIV absorption strength with specific star formation rate (sSFR). In Figure \ref{fig: color ssfr}, the left panels show the variation of \CIV equivalent width (top left) and AOD \CIV column density (bottom left) with the sSFR of the COS-Dwarfs galaxies. We restrict this plot to include lines of sight passing within 0.5{\rvir}  because we only detect \CIV absorption in these sight lines (with a typical detection threshold of $\approx$ 50-100 {m\AA}).  We find that for the 24 galaxies that are identified as star-forming, there are 15 detections of \CIV absorption (detection probability $\rm{ P = 0.63 \pm 0.096}$ at 68\% confidence), while for the 8 passive galaxies there are two \CIV detections (detection probability  $\rm{P = 0.25 \pm 0.14}$ at 68\% confidence). If we exclude the three non-detections with a 3$\sigma$ detection threshold > 100 m{\AA} (all star-forming galaxies), the \CIV detection probability for star-forming galaxies increases to  $\rm{ P = 0.71 \pm 0.1}$ at 68\% confidence interval. This suggests that the probability of detecting \CIV absorption around a star-forming galaxy is generally higher than around a passive galaxy. 

We test the statistical significance of the correlation of \CIV absorption strength with host galaxy sSFR, by performing a generalized Kendall's tau test. We only consider absorber-galaxy pairs within {\rvir} $ \leq $ 0.5. We obtain $\tau$ = 0.35 (significance P = 0.048) and hence can reject the null hypothesis that there is no correlation between \CIV absorption equivalent width and host galaxy sSFR at the 95\% confidence level. We select all the galaxies within {\rvir} $ \leq $ 0.5 and perform a two-sample KS test that rules out the null hypothesis that the star-forming and passive galaxies are drawn from the same parent distribution of equivalent widths at 98\% confidence level (2.3$\sigma$). We interpret these results as tentative, but not conclusive evidence that there is a direct relationship between \CIV absorption strength and sSFR. 

The correlation between galaxy star formation with stellar mass could potentially influence this apparent correlation found above. This can hinder our ability to conclusively identify star formation as the key influence on the presence of \CIV. To assess whether star formation and not simply stellar mass has a direct relationship with the observed \CIV, we select galaxies with $\rm{9 \leq \log M_{*}/M_{\odot}  \leq 10 }$. In this range, there are 22 star-forming and 7 passive galaxies. We perform a two-sample KS test that rejects the null hypothesis that the star-forming and passive galaxies draw from the same parent distribution of equivalent widths at $>$ 99.5\% confidence level (2.8$\sigma$). This result supports our interpretation that there is tentative, but not conclusive evidence for a direct relationship between \CIV absorption strength and sSFR. 

The variation of \CIV absorption strength with sSFR is further shown by adding the sample of \cite{Borthakur2013} (green squares), and the three COS-Halos galaxies for which we have \CIV coverage (Figure \ref{fig: color ssfr}, right panels). The COS-Halos galaxies are marked with open circles for identification. These are primarily {\lstar} galaxies and four out of the five \cite{Borthakur2013} galaxies with detected \CIV absorption are star-burst galaxies at high impact parameters.  Combining the two samples, we find that out of the 33 galaxies that  are identified as star-forming, there are 21 detections of \CIV absorption (detection probability $\rm{ P = 0.64 \pm 0.08}$ at 68\% confidence) and for the 18 passive galaxies there are three \CIV detections (detection probability  $\rm{P = 0.17 \pm 0.09}$ at 68\% confidence).

This suggests that at these equivalent width/column density limits, if a strong ($W_r \gtrsim 100$ m{\AA}) \CIV absorption system is observed, it is more likely to be associated with a star-forming galaxy. One possible explanation for this high \CIV detection rate amongst star-forming galaxies, along with the low covering fraction of \CIV absorption beyond {0.5\rvir},  could be that the \CIV absorption is driven into the CGM by star-formation driven winds \citep{Wiener2009,Rubin2011,Bordoloi2013b} and they are falling back into the galaxy as a galactic fountain with a turn around radius of $\approx$ 0.5\rvir.

\begin{figure}[t!]
\centering
\includegraphics[height=6.cm,width=7cm]{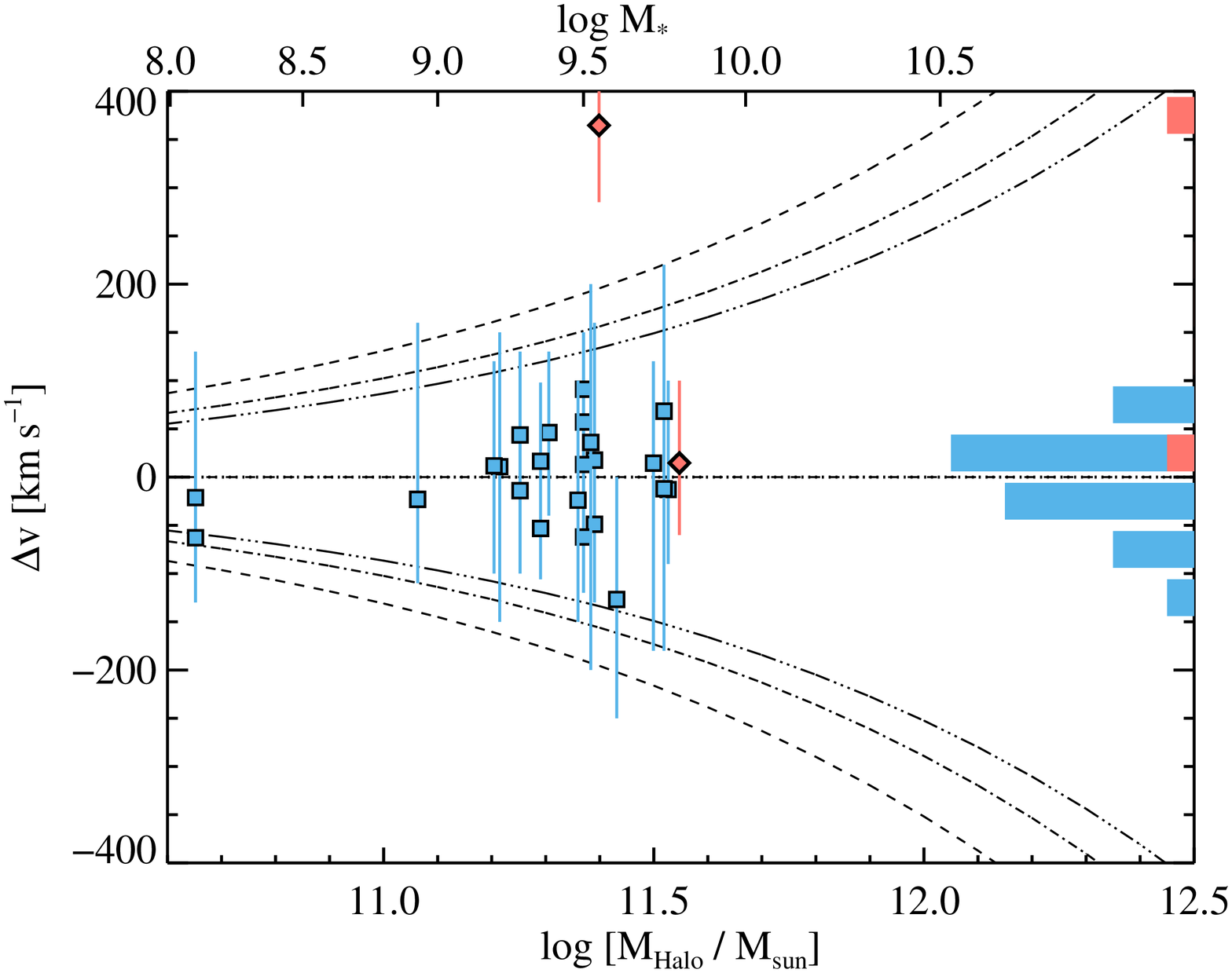}     
    \caption{The \CIV component velocity centroids with respect to the galaxies' systemic redshifts as a function of the inferred dark matter halo mass for star-forming (blue squares) and passive (red diamond) galaxies, respectively. The range bars indicate the maximum projected kinematic extent of \CIV absorption for each system. The histogram represents the distribution of individual component velocities. The dashed lines show the mass-dependent escape velocities at R = 50, 100 and 150 kpc, respectively.}
\label{fig: civ fit}
\end{figure}

\subsection{\CIV Kinematics}

In this section we present the kinematics of the observed \CIV absorption profiles and discuss whether the observed absorption is consistent with being bound to the dark matter halo of the host galaxy. We obtain the \CIV column densities  and kinematics by Voigt profile fitting as described in Section 3.  Figure \ref{fig: civ fit} shows the velocity centroids for each Voigt profile fitted component as a function of the stellar mass and the inferred dark matter halo mass for star-forming (blue squares) and passive (red diamonds) galaxies. The range bars indicate the velocity spread of each \CIV absorption line system that were used to compute the equivalent widths of these systems. These are equivalent to the full width at zero optical depth. The full velocity width includes contributions from both thermal broadening and bulk flow and this is a proxy for the maximum possible (projected) kinematic extent of the absorption. The distribution of the component velocities are clustered around the systemic velocity of the host galaxy, with a median velocity of 13 \kms and a standard deviation of 50 $\rm{kms^{-1}}$. There is one exception, with one system being at $\approx$ 350 {\kms} from its host galaxy. However, statistically most of the detected \CIV absorption is  closely associated with the galaxies in velocity space. This is perhaps not surprising, since we are selecting lines of sight very close to the foreground galaxies. But there is no observational reason or systematic error or selection effect that prevents us from more commonly detecting strong \CIV absorption at high velocities relative to the systemic redshift of the associated galaxies.

Figure \ref{fig: civ fit} also compares these velocities with the escape velocities of the halos in which they reside. We convert the stellar masses of the galaxies to the total dark matter halo mass by using the method described in \cite{Moster2010}. We assume a spherically symmetric NFW profile (with concentration parameter, c=15) and calculate the escape velocity as a function of halo mass at three different radii (R = 50, 100, 150 kpc respectively). In Figure \ref{fig: civ fit} these mass-dependent escape velocities are shown as dashed lines. We see that little of the fitted absorption velocity centroids exceed the estimated escape velocities of these galaxies. Some of the velocity ranges are comparable to the escape velocities (range  bars) but that can be attributed to the line wings. We conclude that most of the detected \CIV absorption is consistent with being bound to the dark matter halos of their host galaxies. 

There is always the chance that these galaxies have associated \CIV absorption at higher relative velocities that remains unseen because it is below our detection limits. We can roughly estimate that any such high velocity absorption components should have a factor of $\approx$ 5-10 lower column densities than the typical detected absorption (Figure \ref{fig: 1d radial profile}, bottom panels), since the detections typically have column densities that are 5-10 times above the detection limits.

\begin{figure*}[htb!]
\centering
    \includegraphics[scale=.315]{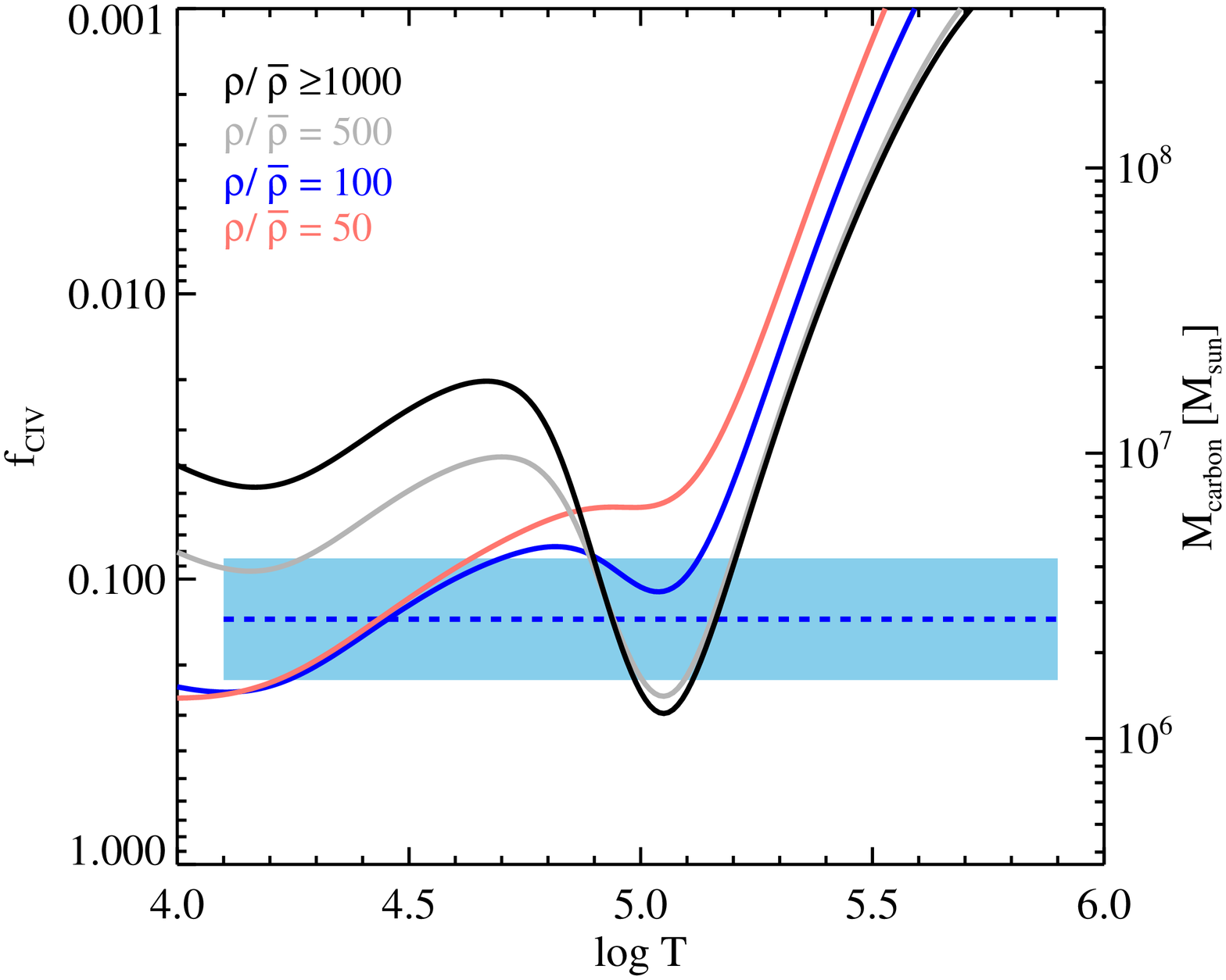} 
    \includegraphics[scale=.315]{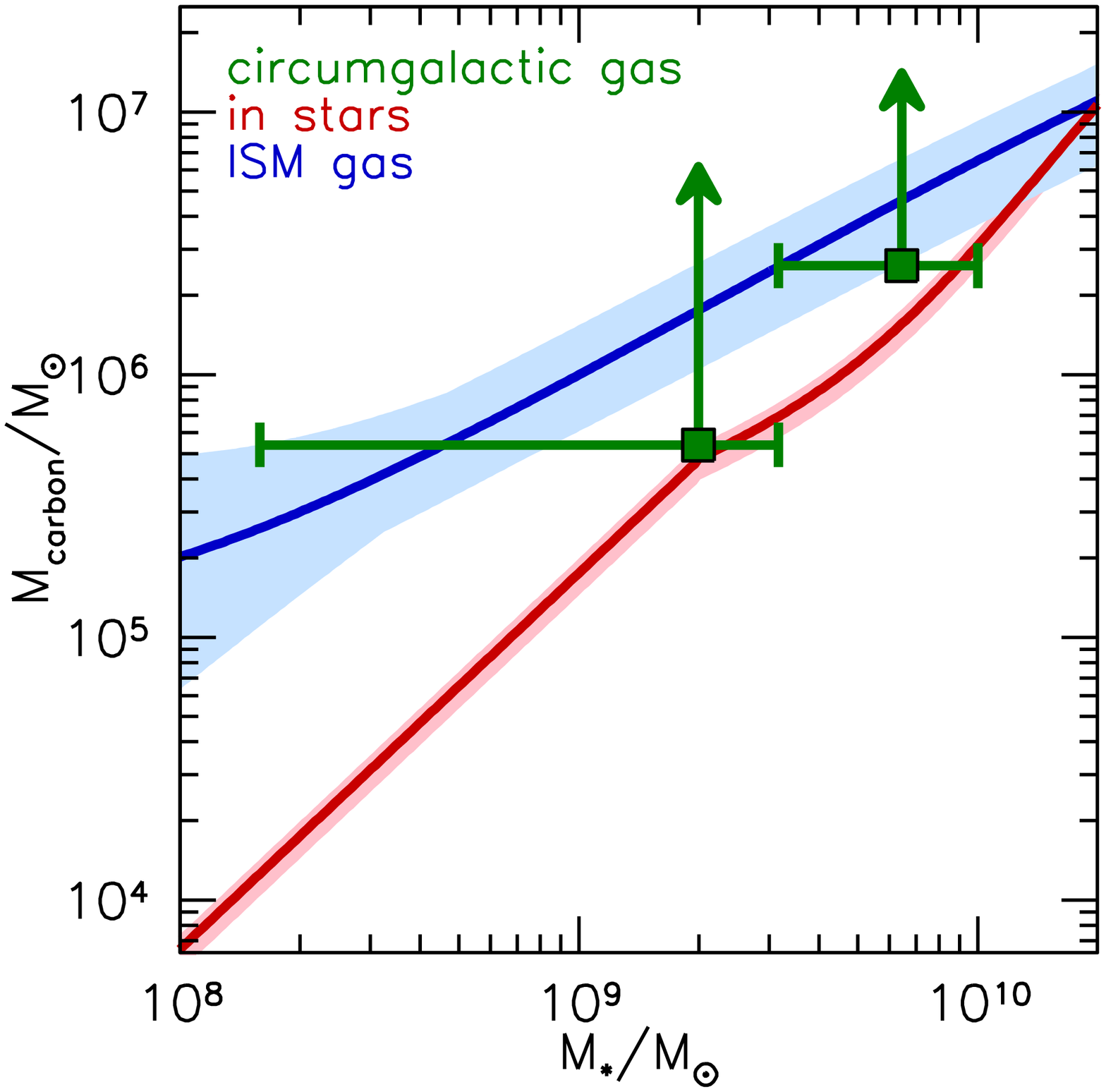}   
  \caption{Carbon mass estimates in the CGM of \subl galaxies compared to their galactic reservoirs. Left Panel: The curves show the variation of the fraction of gas-phase carbon in the \CIV ionization state ($f_{CIV}$) with temperature for four overdensities relative to the cosmic mean ($\rm{\rho/\bar{\rho}}$). For $\rm{\rho/\bar{\rho} \; \geq}$ 1000 (black curve), collisional ionization dominates. For the lower overdensities photoionization by the extragalactic background can increase $f_{CIV}$ at low temperatures. The blue band shows the expected carbon mass in the galaxies' ISM at $\rm{\log M_{*} = 9.5 }$, if the galaxies lie on the standard $M_{ISM}$ vs $M_{*}$ relation and follow the mass-metallicity relation (MZR). Right Panel: The CGM carbon mass (green points) compared to the interstellar carbon mass (blue band) and the carbon mass in stars (red band) as a function $M_{*}$. The green points show the conservative minimum carbon mass for blue galaxies in those mass ranges (Table \ref{mass_table}).}
\label{fig: mass estimate}
\end{figure*}

\section{Minimum mass of CGM carbon}
One of the striking findings of the COS-Halos survey was that around {\lstar} galaxies, there is about as much oxygen (proxied by O VI) as found in the ISM of those galaxies \citep{Tumlinson2011a}. Analysis of the lower ionization transitions further showed the presence of a large reservoir of metals and gas in the cool CGM (conservatively, $\rm{M_{CGM}^{cool} > 10^{9} \; M_{\odot}}$; \citealt{Werk2014}).  This substantial metal budget helps in bridging the missing metals budget around {\lstar} galaxies \citep{Peeples2013a}. The COS-Dwarfs survey is well suited to estimate the  carbon mass in the CGM around {\subl} galaxies. The range of physical conditions in diffuse gas that can contain significant amounts of \CIV is narrow enough that we can obtain robust lower limits on the total mass of carbon in the CGM of dwarf galaxies. 

The total \CIV mass encompassed within an impact parameter (R) is given as,

\begin{equation}
\rm{
M_{CIV} = \pi R^{2} \langle N_{CIV}\rangle  12m_{H} C_{f} ,
}
\end{equation}
where $\rm{C_{f}}$ is the mean covering fraction of \CIV absorption around galaxies within radius R and $\rm{ \langle N_{CIV}\rangle}$ is the mean \CIV column density within R. We obtain a robust lower limit on  carbon mass ($M_{carbon} $) by applying a conservative ionization correction to $M_{CIV}$. To estimate the ionization correction, we have considered the ionization state of \CIV over a wide range of temperatures using the CLOUDY photoionization code \citep{cloudy1998}, assuming ionization equilibrium and including both collisional ionization and photo-ionization similar to \cite{Tumlinson2011a}. Under typical CGM physical conditions, the ionization timescales are of the order of $\sim 10^{7-8}$ years or less, which is much smaller than the halo dynamical timescales of $10^{9}$ years. Hence it is unlikely that a large fraction of the CGM gas would be far from ionization equilibrium. In Figure \ref{fig: mass estimate} (left panel), the curves trace the fraction of gas-phase carbon in the \CIV ionization state ($\rm{f_{CIV}}$) as a function of temperature for four overdensities relative to the cosmic mean density ($\rm{\rho / \bar{\rho}}$). At the highest overdensities ($\rm{\rho / \bar{\rho} \; \geq}$ 1000, black curve), collisional ionization dominates and for all lower overdensities,  photoionization by the extragalactic background would cause the increase of $f_{CIV}$ at low temperatures. The blue band shows the expected carbon mass of the galaxies' ISM if they lie on the standard relation between $M_{ISM}$ and $M_{*}$ and follow the mass-metallicity relation (MZR) at $\rm{\log M_{*} = 9.5 }$. Regardless of whether photoionization or collisional ionization dominates, we assume the most conservative measure of the ionization correction to estimate the carbon mass as ${f_{CIV} = 0.3}$.

Scaling for values typical to our sample and applying a conservative \CIV correction, we find that the minimum carbon mass for all galaxies is
\begin{equation}
\rm{
M_{carbon} \gtrsim 1.2 \times 10^{6} \;M_{\odot} \left(\frac{N_{CIV}}{10^{14}cm^{-2}} \right) \times  \left( \frac{R}{110 \,kpc }\right)^{2}  \times \left( \frac{0.3} {f_{CIV}} \right),
}
\label{mass_eqn}
\end{equation}
where we find that within 110 kpc, 17 out of 32 galaxies show \CIV absorption. The covering fraction and mean column densities are measured in two radial bins: 8 out of 11 detections within $R < 50$ kpc and 9 out of 21 detections within $50 \leq R \leq 110$ kpc. The column densities used in equation \ref{mass_eqn} are derived from Voigt profile fitting and represent the entire range in stellar mass. The mean column density within $R < 50$ kpc is $\rm{4 \times 10^{14}\; cm^{-2} }$ and within $50 \leq R \leq 110$ kpc is $\rm{8.2 \times 10^{13}\; cm^{-2} }$. We estimate $M_{carbon}$ using both Voigt profile fitting and AOD column densities. Carbon masses derived using both methods are tabulated in Table \ref{mass_table}. The column densities derived using the AOD method are lower limits for saturated lines; therefore the carbon masses are only conservative lower limits. The column densities derived from Voigt profile fitting are more reliable, however for severely saturated lines, they are also probably underestimating the column densities. 

If these galaxies lie on the stellar metallicity relation \citep{woo08}, the mean trend of gas fractions, and the gas-phase mass-metallicity relation \citep{Peeples2013a} for low-z galaxies, then they have interstellar carbon masses of $M^{C}_{ISM} = 1.6\times 10^{5}$ to $5\times 10^{6}$ {\msun} and masses of carbon in stars ($M^{C}_{star}) = 10^{4}$ to $1.5 \times 10^{6}$ {\msun}, where we have assumed solar abundance ratios of C/Z=0.18 by mass (i.e., $[12+\log(C/H)]_\odot=8.50$ and $Z_\odot=0.0153$; \citealt{caffau11}). Figure \ref{fig: mass estimate} (right panel), shows the minimum CGM carbon mass of star-forming galaxies (green points), for $f_{CIV} =0.3$, compared with the interstellar carbon mass (blue band) and stellar carbon mass (red band) as a function of mass \citep{Peeples2013a}.  The range in the blue band denotes the uncertainty in the ISM carbon mass owing primarily to uncertainties in the calibration of the gas-phase metallicity indicators and partially to the uncertainties in the gas fractions. The narrower red band denotes systematic uncertainties introduced to the mass of carbon in stars from uncertainties in the overall solar abundance scale but does not take into account possible non-solar abundance ratios. Our key finding is that the minimum CGM carbon mass is 50\% to 80\% of the total ISM carbon mass and is always higher than the total mass of carbon in stars.

The ratio of carbon mass in the ISM to the CGM remains more or less constant for the two different mass bins probed (assuming a constant ionization correction). It should be stressed that the CGM carbon masses quoted here are conservative lower limits, as we assume a limiting ionization correction of $f_{CIV} =$ 0.3 and the \CIV column densities may be underestimated owing to saturation. For the typical densities expected at R $\approx$ 100 kpc,  $f_{CIV}$ exceeds 0.3 only at T $\approx\; 10^{5.05}$K and $f_{CIV}$ exceeds 0.05 at T $\approx\; 10^{4.8} - 10^{5.25}$K.  Hence the total carbon mass can easily be even higher by a factor of 6, higher than the ISM carbon mass in these galaxies. This result is analogous to the findings of COS-Halos oxygen metal budget \citep{Tumlinson2011a}, where it was found that a significant mass of oxygen exists in the CGM of \lstar galaxies, and is comparable to the mass of oxygen in the ISM of  those  galaxies.

We estimate the carbon mass out to 110 kpc, beyond which no \CIV absorption is detected. However, it is plausible that some diffuse \CIV gas is present below the survey detection threshold, at larger radii. To ascertain how much carbon could be ``hidden'' beyond 110 kpc, we extend equation \ref{mass_eqn} out to $\approx$ 220 kpc. We assume  a mean column density at $R > 110$ kpc of $\rm{3 \times 10^{13}\; cm^{-2} }$, which is comparable to our typical detection limits. We compute the total carbon within 220 kpc and find that the total increase in carbon mass in going from 110 kpc to 220 kpc is $\approx$ 20\%. Hence the bulk of the carbon mass observed in the CGM should be observed within 110 kpc (which corresponds to roughly 0.5 \rvir) of the host galaxy.   

\begin{figure}
\includegraphics[height=7.5cm,width=8.5cm]{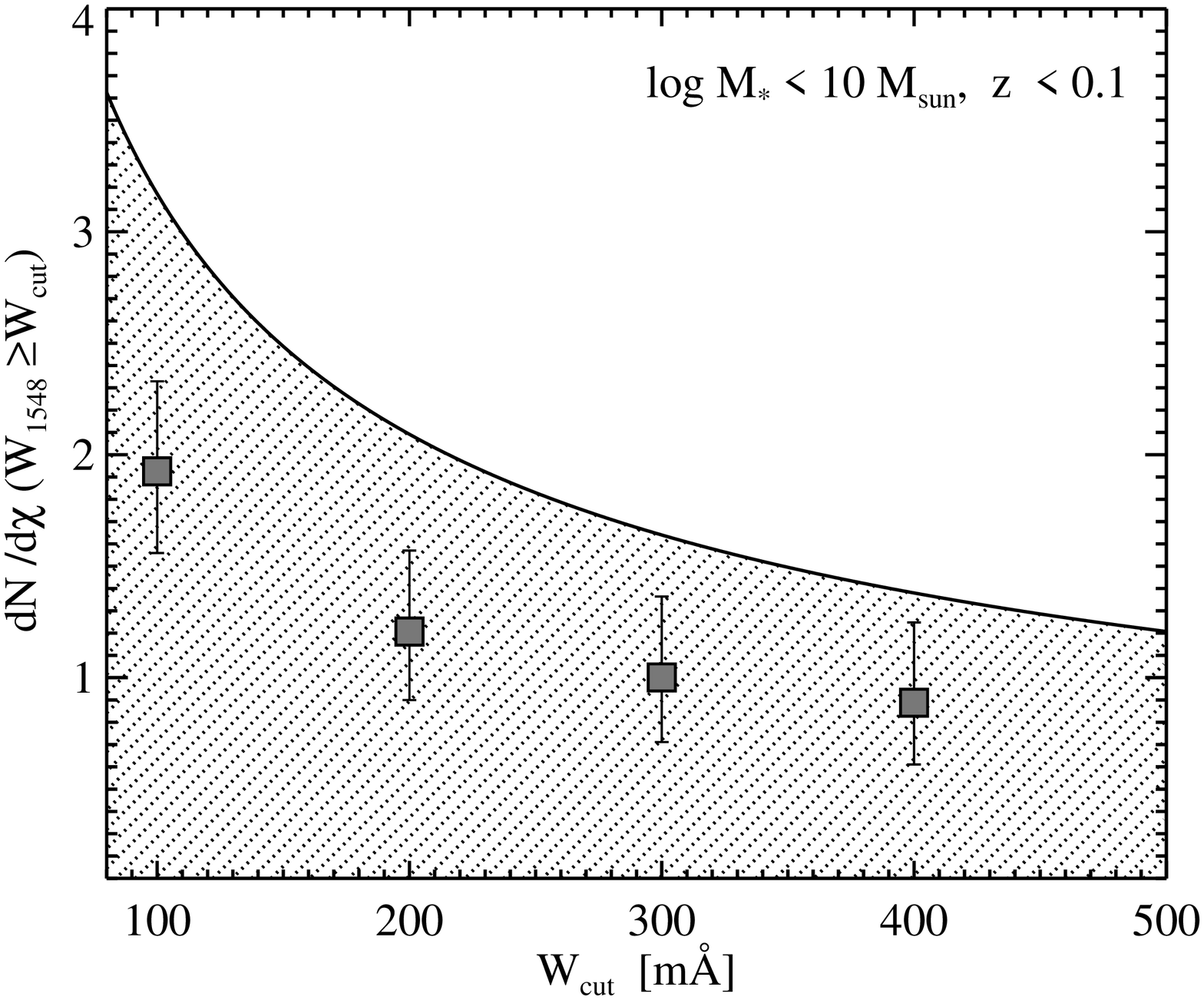}
\caption{Cumulative absorber line density as a function of \civ equivalent width. The summed $dN/d\chi$ for COS-Dwarfs galaxies are shown as gray squares. The hashed region shows the cumulative $dN/d\chi$ at z$<$0.4, adopted from \cite{Cooksey2010}. COS-Dwarfs galaxies can account for $\sim$ 60\% of all low-z \CIV absorbers.}
\label{fig:incidence}
\end{figure}

\section{ Expected incidence of \CIV absorbers}
In this section we estimate the expected redshift-path incidence of \CIV absorbers exceeding a particular absorption threshold $W_{cut}$, around $\log M_* \; \leq \; 10^{10}$\msun galaxies at $z \approx 0$. This exercise is the opposite of one undertaken for blind QSO absorption line studies (e.g. \citealt{Cooksey2010}, \citealt{Shull2014}); where one takes the observed number of absorption systems per unit redshift and calculates the CGM cross section and number density required to account for the rate of incidence, assuming that the cross-section is all contributed by galaxies. \cite{TumlinsonFang2005}  used the observed number density distributions of intergalactic \OVI absorbers to constrain the  metal distribution in the low-$z$ IGM, and found that observed dN/dz of \OVI absorbers can be explained by the extended CGM of \subl galaxies. \cite{Prochaska2011c} performed a similar calculation based on their analysis of \OVI surrounding z $\sim$ 0.1 galaxies and concluded that the majority of \OVI systems are associated with the extended CGM of \subl galaxies.

Here we infer the rate of incidence of \CIV absorption around the COS-Dwarfs sample exceeding a particular absorption threshold $W_{cut}$ defined as,
\begin{equation}
\rm{
\frac{dN}{d\chi} (W > W_{cut}) = \frac{c}{H_{0}} C_{f}\; n_{gal}\; \pi R_{max}^2
}
\end{equation}
where $\rm{C_{f}}$ is the covering fraction of absorbers with $\rm{(W > W_{cut})}$ and $\rm{R_{max}}$ is the maximum impact parameter at which  an $\rm{(W > W_{cut})}$ absorber is observed. For COS-Dwarfs, we have a sample of galaxies at $z <$ 0.1 with a well defined stellar mass function \citep{Moustakas2013}. We integrate the double Schechter function fit to the  $z <$ 0.1 mass distribution within $8 \leq \log M_*/M_{\odot} \leq 10$ and get a cumulative number density of COS-Dwarfs like galaxies $\rm{n_{gal} \approx 10^{-1.39} }$ Mpc$\rm{^{-3}}$. These values translate to a rate of incidence at $W > 100$ {m\AA}, $dN/d\chi \approx \; 1.93 \pm 0.4$, at $W > 300$ {m\AA}, $dN/d\chi \approx \; 1.0 \pm 0.32$ and at $W > 400$ {m\AA}, $dN/d\chi \approx \; 0.89 \pm 0.36$. The observed incidence of \CIV systems at the same equivalent width threshold at z$<$0.4 is reported to be for $(W > 100$ {m\AA}) $\approx $ 3.17,  for $(W > 300$ {m\AA}) $\approx $ 1.64 and for $(W > 400$ {m\AA}) $\approx $ 1.38 \citep{Cooksey2010}. We present the cumulative absorber line density as a function of \CIV equivalent width for COS-Dwarfs in Figure \ref{fig:incidence}. We adopt the \CIV frequency distribution function at $z <$0.4, described in \cite{Cooksey2010}, and it is integrated to infer the absorber line densities at different limiting equivalent widths. This is shown as the hashed region in Figure \ref{fig:incidence}. It should be noted that all the line densities (and the corresponding error estimates) shown in Figure \ref{fig:incidence} are correlated. Comparing the line densities measured for COS-Dwarfs (gray squares) with the total cumulative absorber line densities (hashed regions) we argue that the low-mass galaxies having $8 \leq \log M_*/M_{\odot} \leq 10$ at z$<$ 0.1, could easily account for almost 61\% (W > 100 {m\AA}, W > 300 {m\AA}), and 64\% (W > 400 {m\AA}) of the strong \CIV absorbers observed today. The remaining $\sim$ 30 to 40\% of the \CIV absorbers can be accounted for by the $>$ \lstar galaxies with a covering fraction of $\approx$ 40\% with $R_{max} \approx$ 200 kpc. It is plausible that 1/2 - 2/3 of strong \CIV in blind IGM samples arise in {\subl} halos.

\begin{figure*}

  \includegraphics[height=5.cm,width=5.85cm,angle=0]{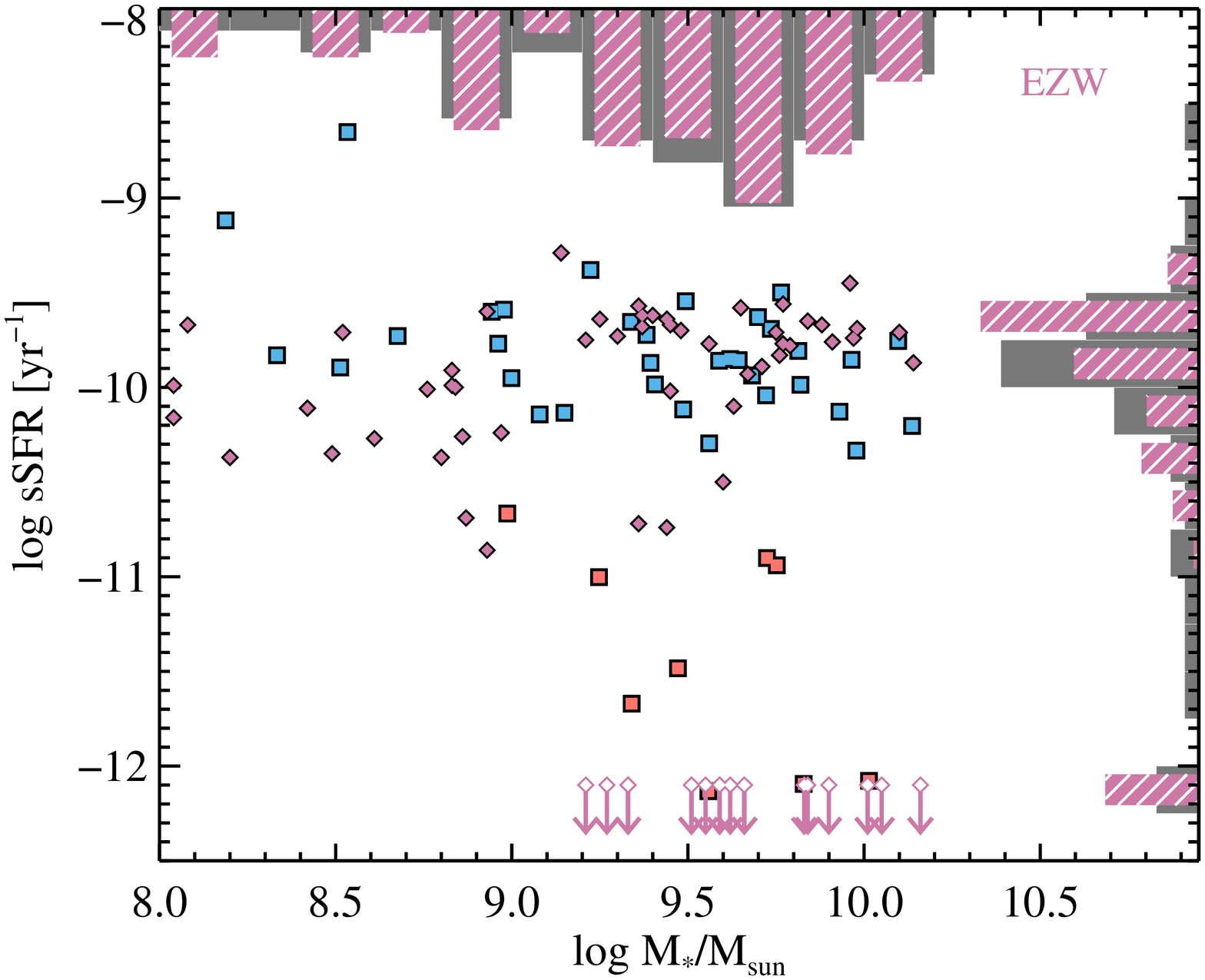}
  \includegraphics[height=5.cm,width=5.85cm,angle=0]{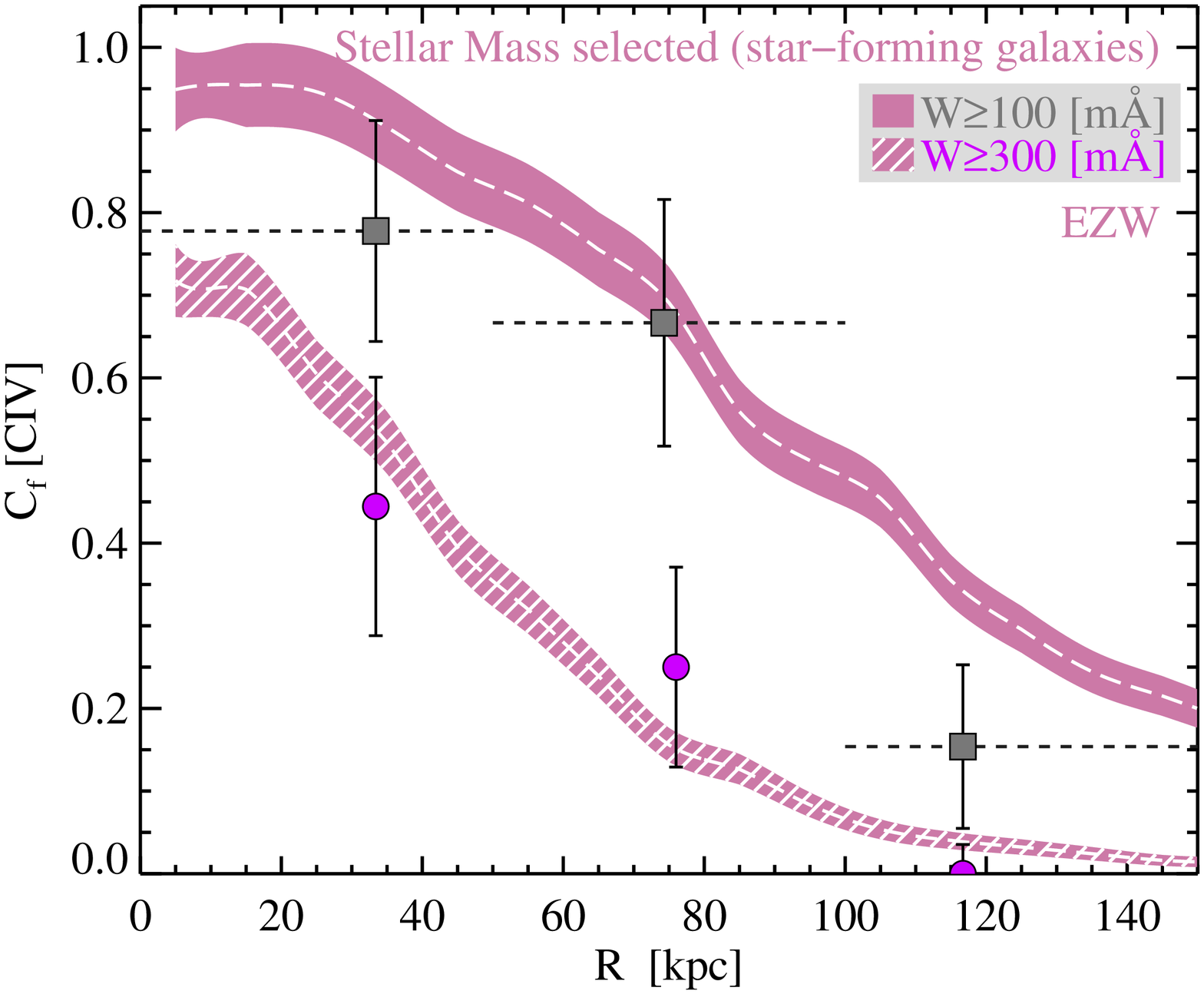}
  \includegraphics[height=5.cm,width=5.85cm,angle=0]{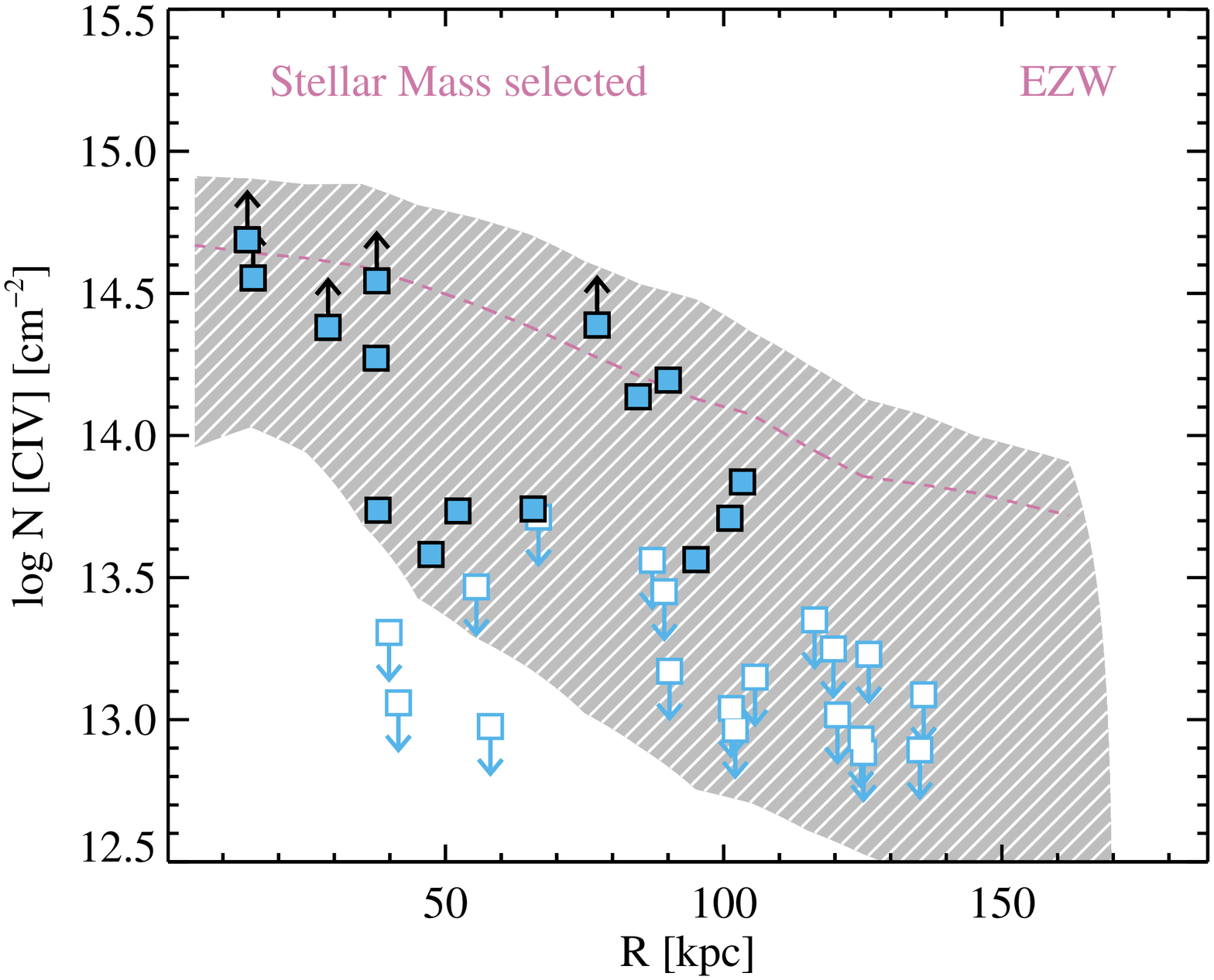}

  \includegraphics[height=5.cm,width=5.85cm,angle=0]{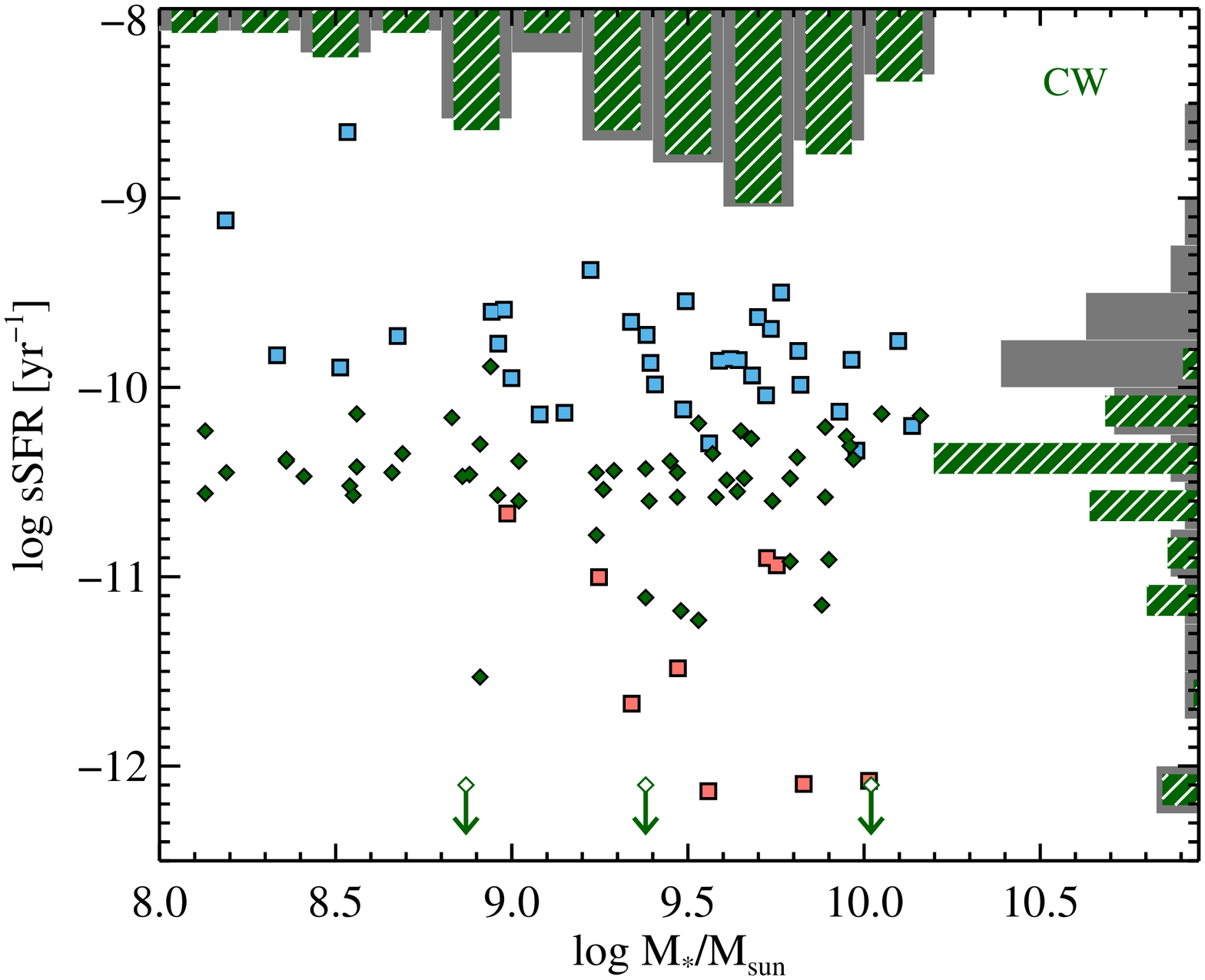}
    \includegraphics[height=5.cm,width=5.85cm,angle=0]{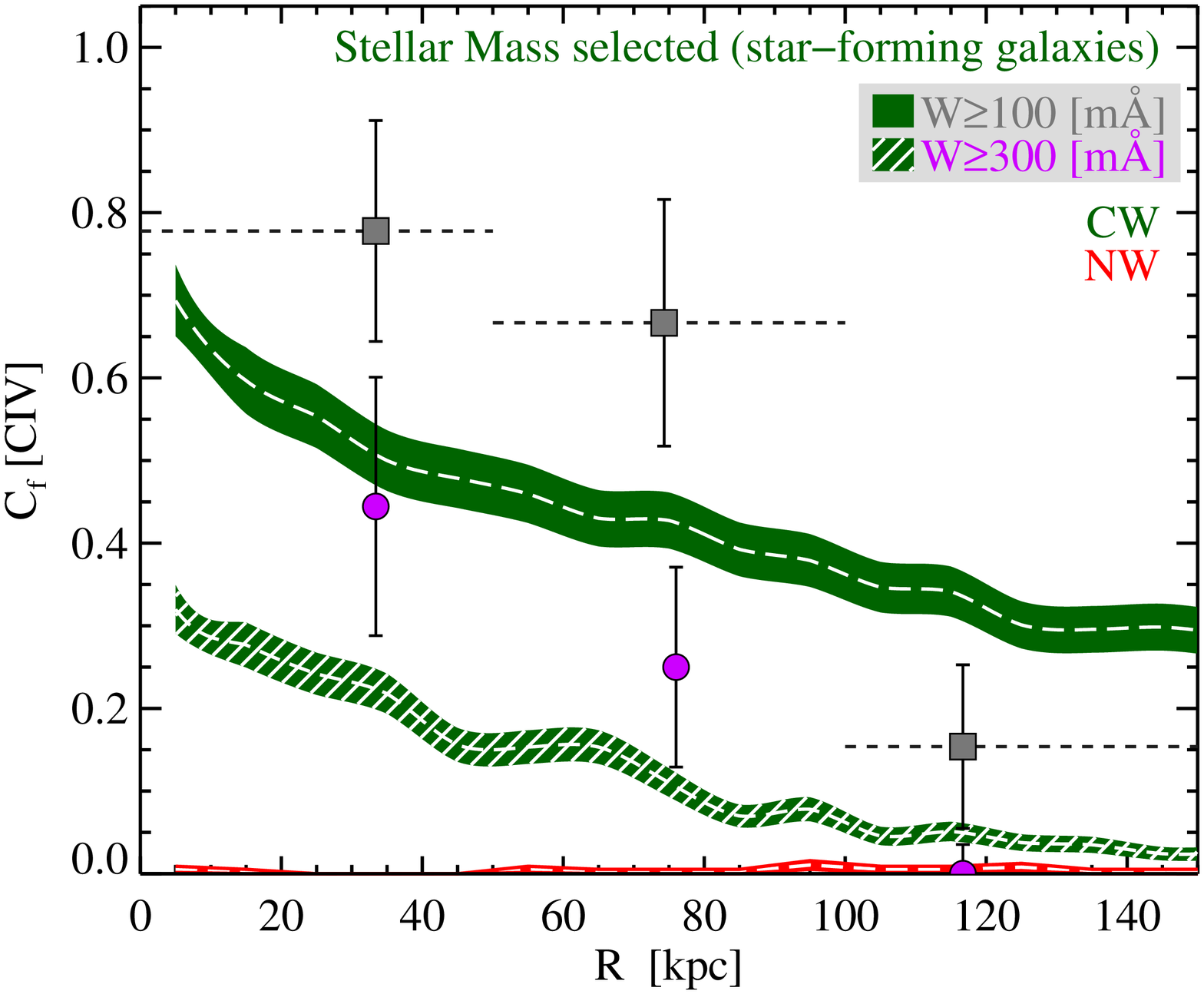}
  \includegraphics[height=5.cm,width=5.85cm,angle=0]{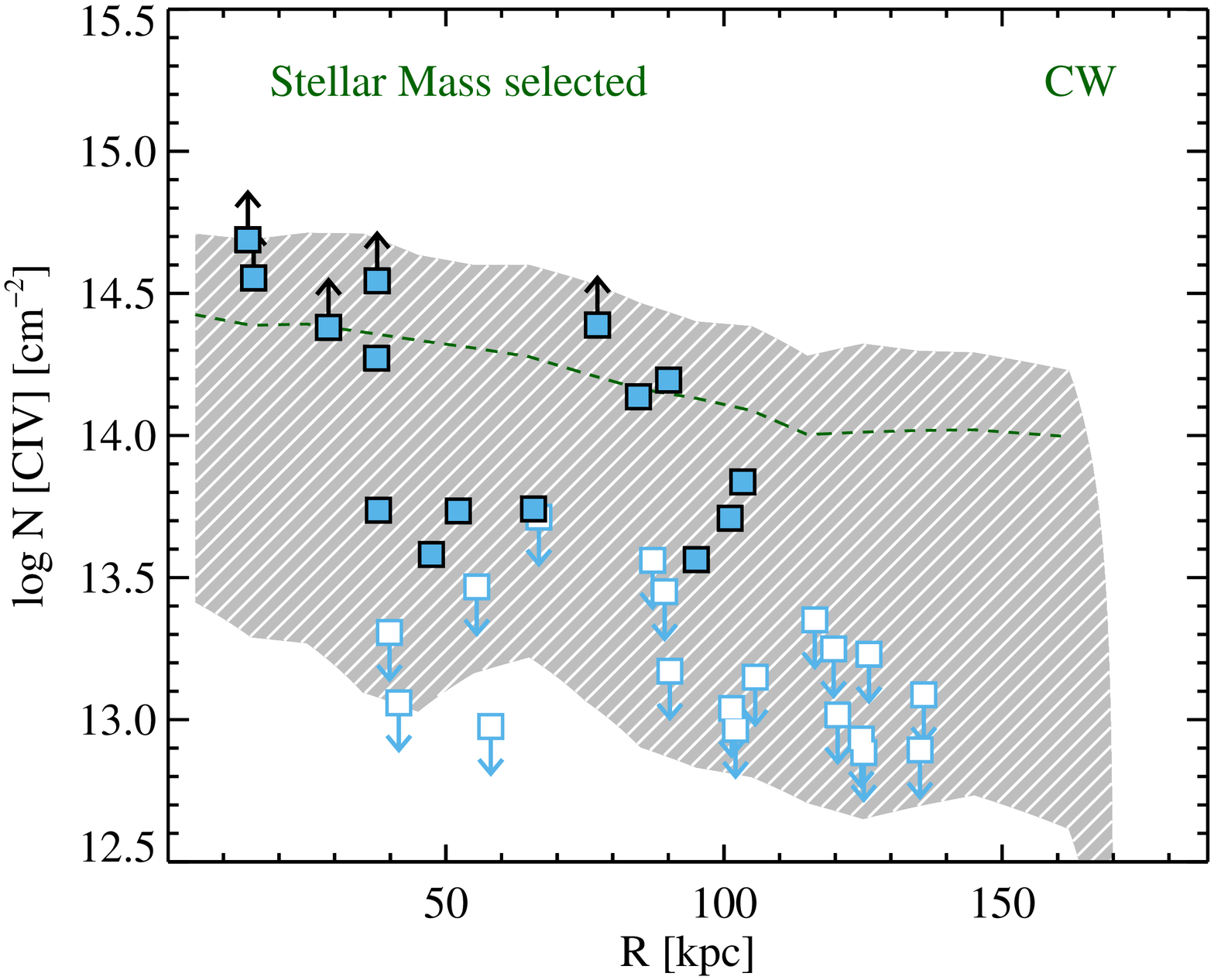}

 \caption{Comparison of different feedback prescriptions in hydro simulations with observations. The distribution of sSFR and stellar mass (left panels) for 117 galaxies with different feedback prescriptions (diamonds) and 43 COS-Dwarfs galaxies (blue and red squares) are shown respectively. The \CIV covering fraction estimates for star-forming galaxies are shown in the middle panels. The purple and green bands represent covering fraction estimates for the two wind models and the red band represents same for the no wind model. The error bars represent the 68\% confidence intervals. Right Panels: The \CIV column density radial profiles of the star-forming galaxies is compared to the simulations. The hashed regions represent the 1$\sigma$ spread of column density in the model sight lines. The dashed green and purple lines indicate the mean column density radial profiles in the simulations.}
\label{fig: simulation_compare}
\end{figure*}

\section{Comparison with Simulations }

We use a set of cosmological \gad~N-body+smoothed particle
hydrodynamical (SPH) simulations \citep{spr05} to compare with our observations of {\CIV}
covering fractions. The simulations have a periodic volume with a box
length of $32\hmpc$, contain $2\times 512^3$ dark matter and gas particles,
and use a $\Lambda$CDM cosmology based on the WMAP 9-year results
\citep{hin13}. The gas particle mass is $4.5\times 10^6 \;$\msun, the
dark particle mass is $2.3\times 10^7\; $\msun, and the softening
length is $\epsilon=1.25\hkpc$.

We run mock sight lines through three simulations exploring different
prescriptions for galactic super-winds around galaxies chosen to match
the stellar masses in passive and star-forming subsamples of
COS-Dwarfs. These simulations were introduced by \citet{dav13},
including the constant wind (cw), the energy-driven wind (ezw) and the no winds (NW)  models. These wind models are defined by parameter choices of $\eta$,
the mass loading factor of star formation-driven winds defined as
$\eta \equiv \dot M_{\rm wind}/\dot M_{\rm SF}$, and wind velocity,
$v_{\rm wind}$. The cw model uses a constant value of $\eta=2$ and
$\vw=680\; \kms$ representing a 100\% conversion efficiency of supernova
energy from stars above $8 M_{\odot}$ to kinetic outflows. The ezw model uses energy-driven relations $\eta\sim \sigma^{-2}$ and $\vw
\sim \sigma$ below $\sigma = 75\; \kms$ and $\eta\sim \sigma^{-1}$ for
higher $\sigma$. A quenching prescription described in \citet{dav13} is used in the ezw
model, but does not affect the galaxy masses we explore for
COS-Dwarfs. The no winds model has no outflows (i.e. $\eta=0$).

The spectral generator {\tt specexbin} casts sight lines at impact
parameters out to 150 kpc around galaxies selected in each $z=0.025$
simulation output to match the COS-Dwarfs sample. The \citet{haa01}
ionization background is assumed but our results are indistinguishable
if \citet{haa12} is used instead.  We search around simulated galaxies
matching the $M_*$ distribution of COS-Dwarfs galaxies divided into
star-forming and passive samples based on the subdivision of sSFR$ =
10^{-10.6} {\rm yr}^{-1}$. This sample results in different
distributions of sSFR for each wind model compared to what is observed
in COS-Dwarfs, with ezw providing the best match, because this wind
model agrees best with the abundance matching (AM) constraints of \citet{beh13a}.  In each
case, we select isolated galaxies defined as not having another galaxy 
$\geq 50\%$ as massive of the targeted galaxy's mass within 300
kpc.

In the $M_*$/sSFR-selected sample, satellite galaxies fulfilling the
isolation criteria are allowed in the sample, but make up only 4-9\%
of the simulated samples. Satellite galaxies are most often selected
in the passive subsample around groups where $M_{\rm halo}>10^{13}
M_{\odot}$, because low-sSFR isolated galaxies are rare in the
simulations.  We cast 4 sight lines at 15 equally-spaced impact
parameters ranging from 5 to 145 kpc around three times as many
galaxies as COS-Dwarfs, or 117 in total, for a total of 7020 mock
sightlines per wind model.  We then sum up the $\CIV$
1548\AA~equivalent width within $\pm 600\; \kms$ of the velocity of the
galaxy.

Figure \ref{fig: simulation_compare} shows the results of the
COS-Dwarfs mock surveys compared to the actual observations. The left
panels show the simulated galaxy parameters of $M_*$ and sSFR for each
wind model: ezw (purple, top panels) and cw (green, bottom panels)
with binned histograms, both compared to the COS-Dwarfs galaxies (gray
histograms, red/blue squares).  The simulated galaxies are required to
have the COS-Dwarfs $M_*$ distribution within bins of 0.2 dex, but the 
simulated sSFR is constrained only to match the star-forming and
passive samples using the division of $10^{-10.6} \;{\rm yr}^{-1}$.  We
note that ezw provides the best fit to the distributions of sSFR.  The
ezw feedback parameters were intentionally tuned to match the AM
constraints of \citet{beh13a} via the selection of the $\sigma=75\kms$
threshold below which energy-driven scalings take effect for ezw.  In
contrast, cw winds are more efficient at suppressing star formation,
and predict a comparatively constant $M_*/M_{\rm halo}$ ratio with
many more massive halos hosting centrals than any other wind model.

The middle panels show the simulated covering fractions of 100
m\AA~and 300 m\AA~{\CIV} as a function of impact parameter
for each wind model, compared to the covering fractions from Section 4.2.
Based on \CIV covering fractions, ezw provides the best overall fits,
although the ezw model over-estimates the extent of 100 m\AA~\CIV
detections (solid bands) at $>100$ kpc. The ezw model better
reproduces the \CIV covering fraction for 300 m\AA~\CIV detections
(hashed band). The right panels show the \CIV column density radial profiles of the
star-forming galaxies as compared to the simulations. The gray hashed
regions represent the 1$\sigma$ spread of column density in the model
sightlines.

The cw model performs poorly in reproducing both CGM and galaxy
properties. At close impact parameters, the cw model systematically
under predicts the \CIV covering fractions and at high impact parameters, it over-predicts the extent of 100 m\AA~\CIV covering fraction (solid band). Previous work has shown
that this model predicts a stellar mass function that is too steep
\citep{Oppenheimer2010} and a present-day sSFR distribution that is
too low for low-mass galaxies \citep{dav11b}. A faster $\vw$ arising
from low-mass galaxies in cw relative to ezw results in a flatter
dependence of $\CIV$ covering fractions as metals are more likely to
be pushed to larger distances.

We compare the sSFR distributions of the ezw and the cw models with
that of the COS-Dwarfs galaxies and find that for the ezw model, a two
sample KS test cannot rule out the null hypothesis that the two sSFR
distributions were drawn from the same parent sample at $>$10\%
significance. For the cw model a two sample KS test rules out the null
hypothesis that the two sSFR distributions are drawn from the same
parent sample at 0.001\% significance level. Further, we perform a
likelihood ratio test to compare which model (ezw or cw) best
represents the observed \CIV covering fractions. We obtain a P value
of $\approx$ 0.01, which indicates that there is strong evidence 
that the ezw model represents the data better than the cw
model. Hence, constraints from both the sSFR distribution and the \CIV
covering fraction suggest that the ezw model better represents the
observations as compared to the cw model.

We further compare the observed \CIV covering fraction with that measured in simulations with no winds (NW) (Figure \ref{fig: simulation_compare}, red band middle panel). The NW model predicts \CIV covering fractions of $\sim$ 1\% to 2\% at all impact parameters. Hence the metallic content of the  CGM around \subl galaxies cannot be explained by tidal debris or ram-pressure stripping alone, as these are the only processes for distributing metals into the CGM in our NW simulations. In a recent study, \cite{JiaLiang2014} also found that metals around low-mass galaxies are primarily concentrated within the inner virial radii of the galaxies.  While they find similar observational trends within their sample, they  conclude that winds are inefficient at these masses, but our quantitative comparison to hydrodynamic simulations with and without winds strongly indicates that strong outflows are necessary explain the observed \CIV in the CGM.

In summary, our exploration of simulations yields the best fits for an ezw
model that ejects cool gas ($T\sim 10^4$ K) at moderate velocities
($\vw=150-300\; \kms$) and high mass-loading factors ($\eta=5-15$).
These winds enrich the local CGM where this metal-enriched gas can
re-accrete back onto the galaxy and sustain the observed $z\sim 0$
sSFR distribution of low-mass galaxies.

Finally, if we select simulated central galaxies based on halo masses
derived from the abundance matching applied to COS-Dwarfs, we would
find more distinguishing power using \CIV covering fractions between
the various wind models, but at the cost of selecting distributions of
$M_*$ and sSFR that do not match COS-Dwarfs.  The ezw model does not
change much because these galaxies agree well with abundance matching constraints,
while cw would have far lower covering fractions and no detections of
300 m\AA~$\CIV$ absorbers, because the galaxies form less stars and
the $\vw=680\; \kms$ heat the CGM, suppressing star formation.  These
trends show that in the low-mass galaxy regime explored by COS-Dwarfs,
our simulations find  that covering fractions scale with galaxy $M_*$.  

\section{Comparison with previous studies}
In this section, we present the combined measurements of previous studies from the literature, which characterized the \CIV absorption profile around galaxies. We stress that this comparison involves galaxies with heterogeneous mass, SFR, and selection, so this is not a statistically rigorous comparison. 

\begin{figure*}[htb!]
\centering
    \includegraphics[height=6cm,width=7cm]{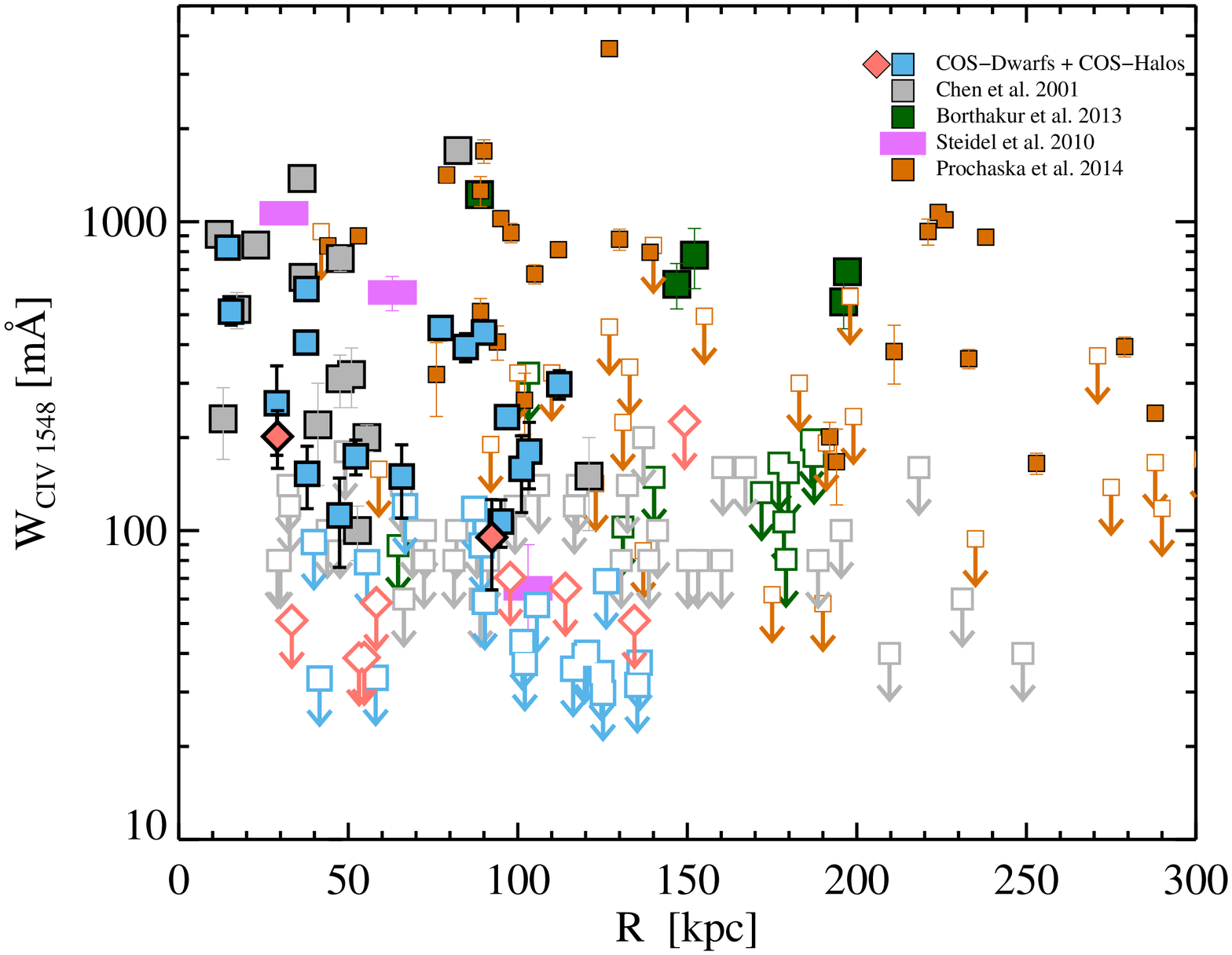}  
    \includegraphics[height=6cm,width=7cm]{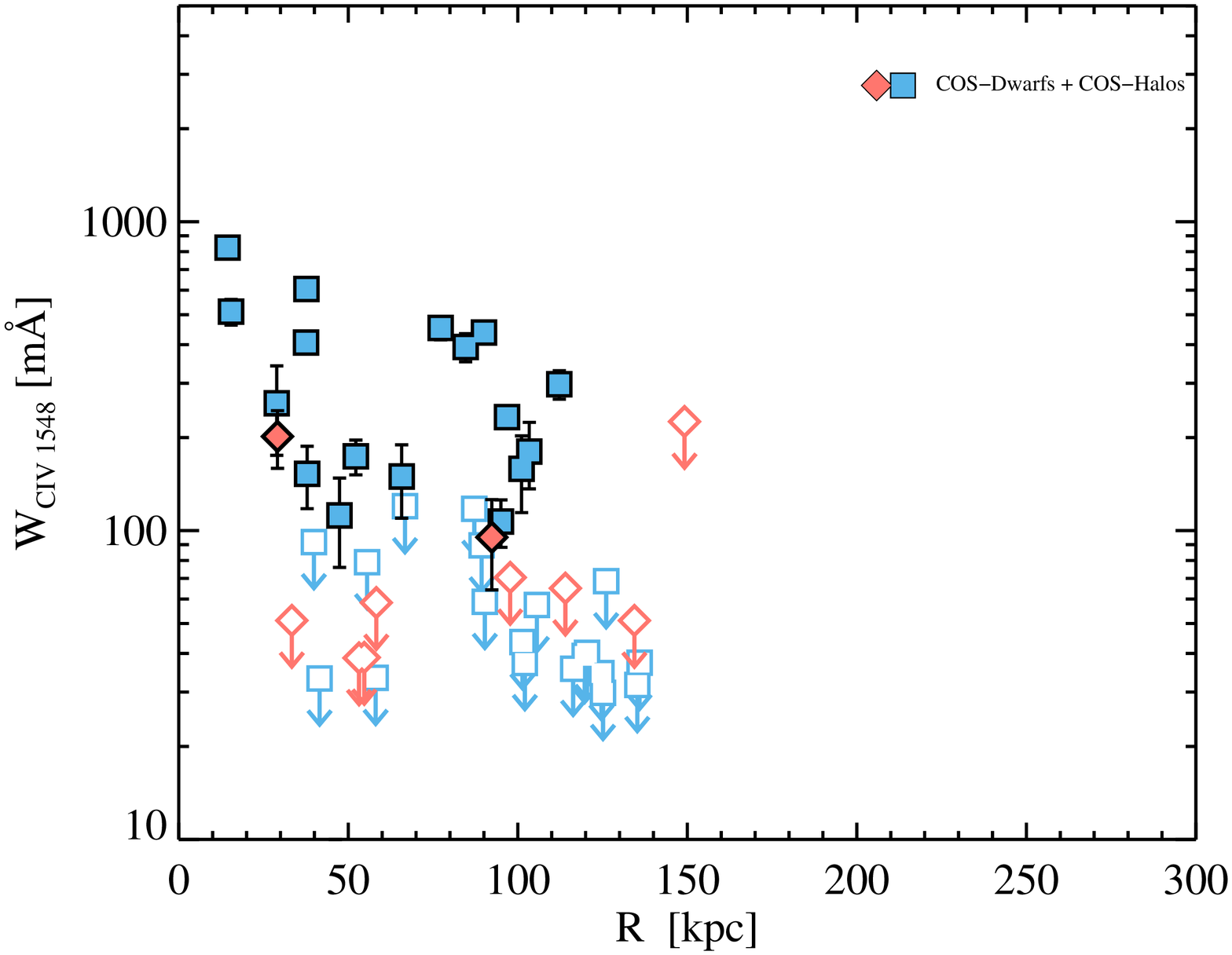}   
    \includegraphics[height=6cm,width=7cm]{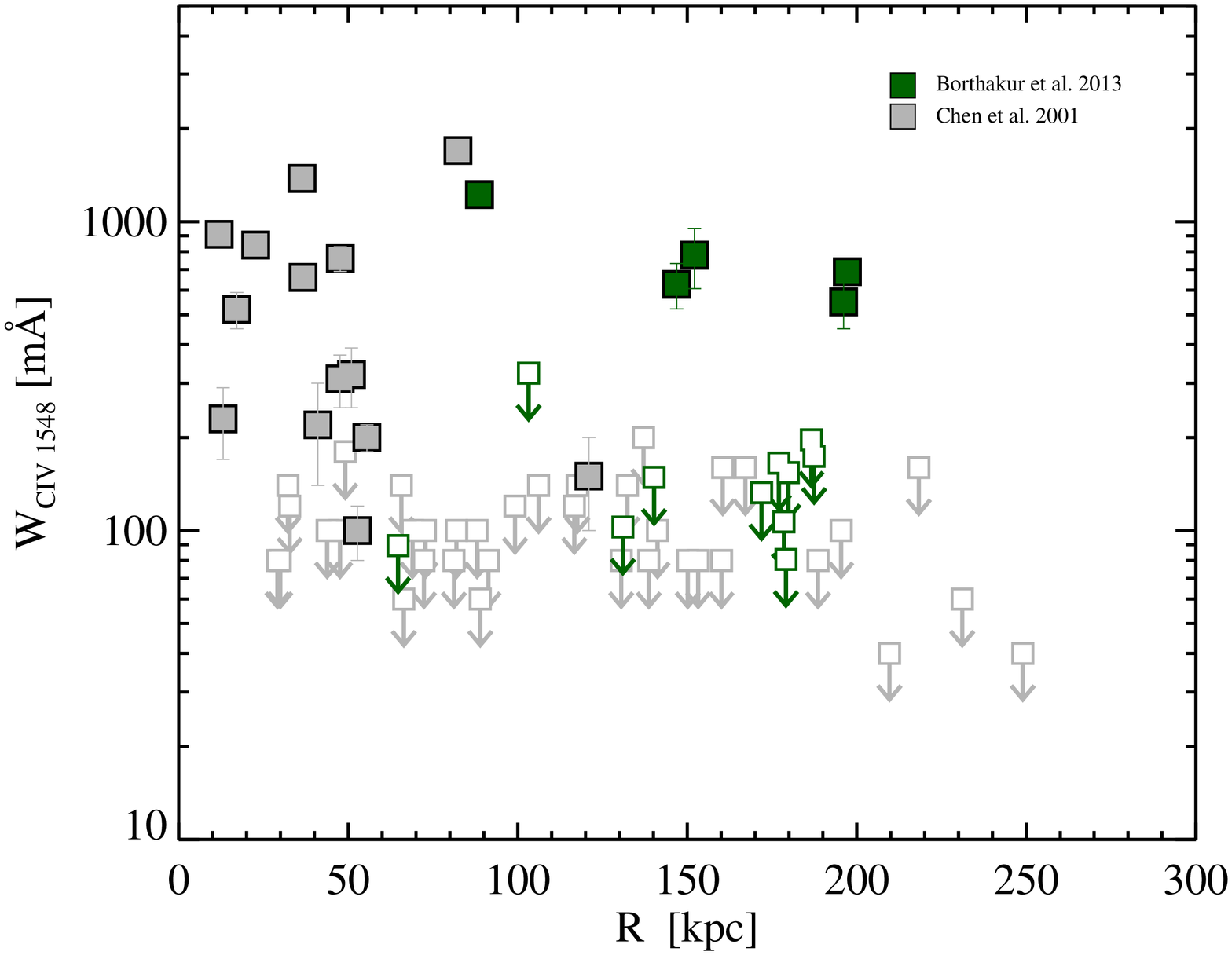}   
    \includegraphics[height=6cm,width=7cm]{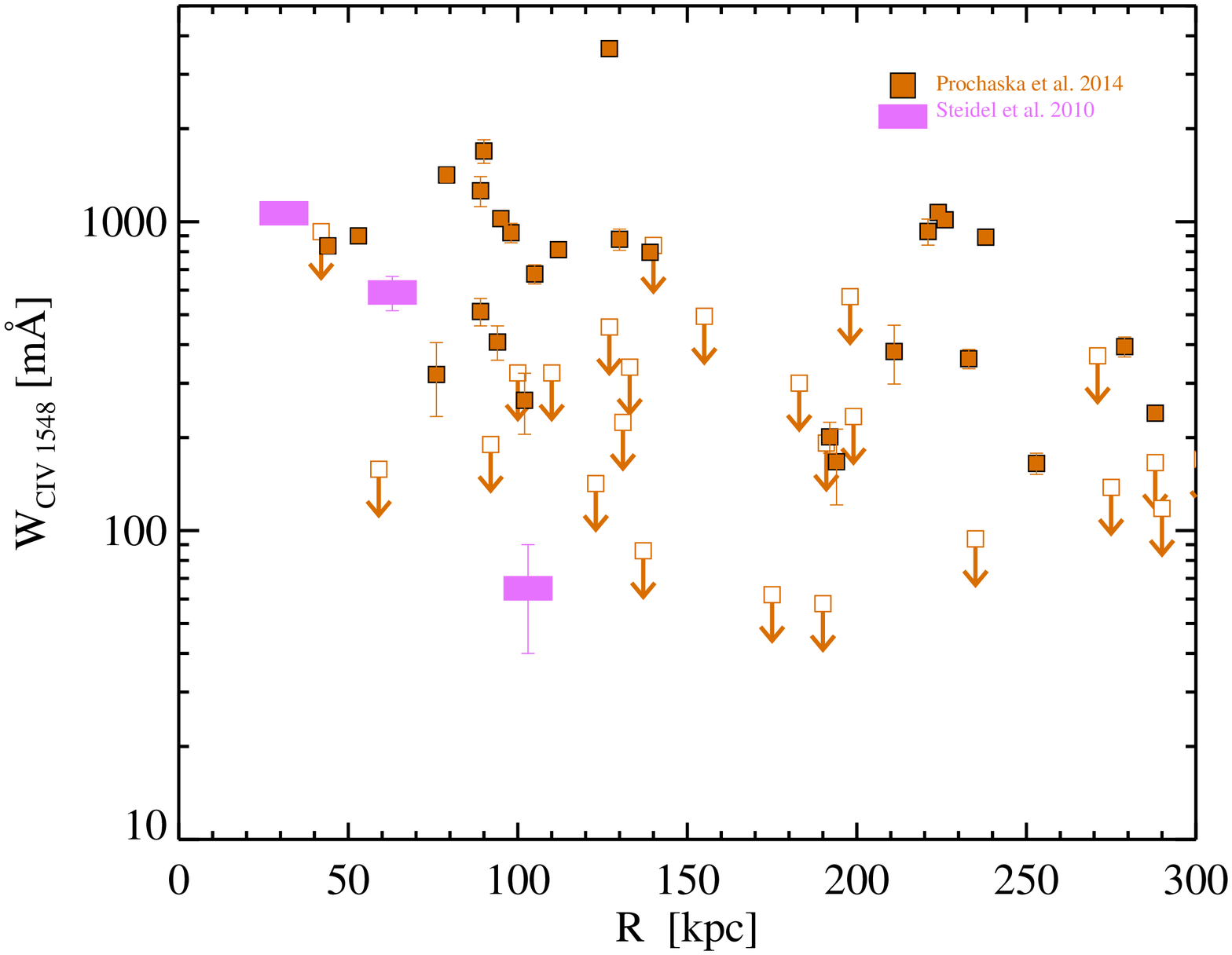}
    \includegraphics[height=6cm,width=7cm]{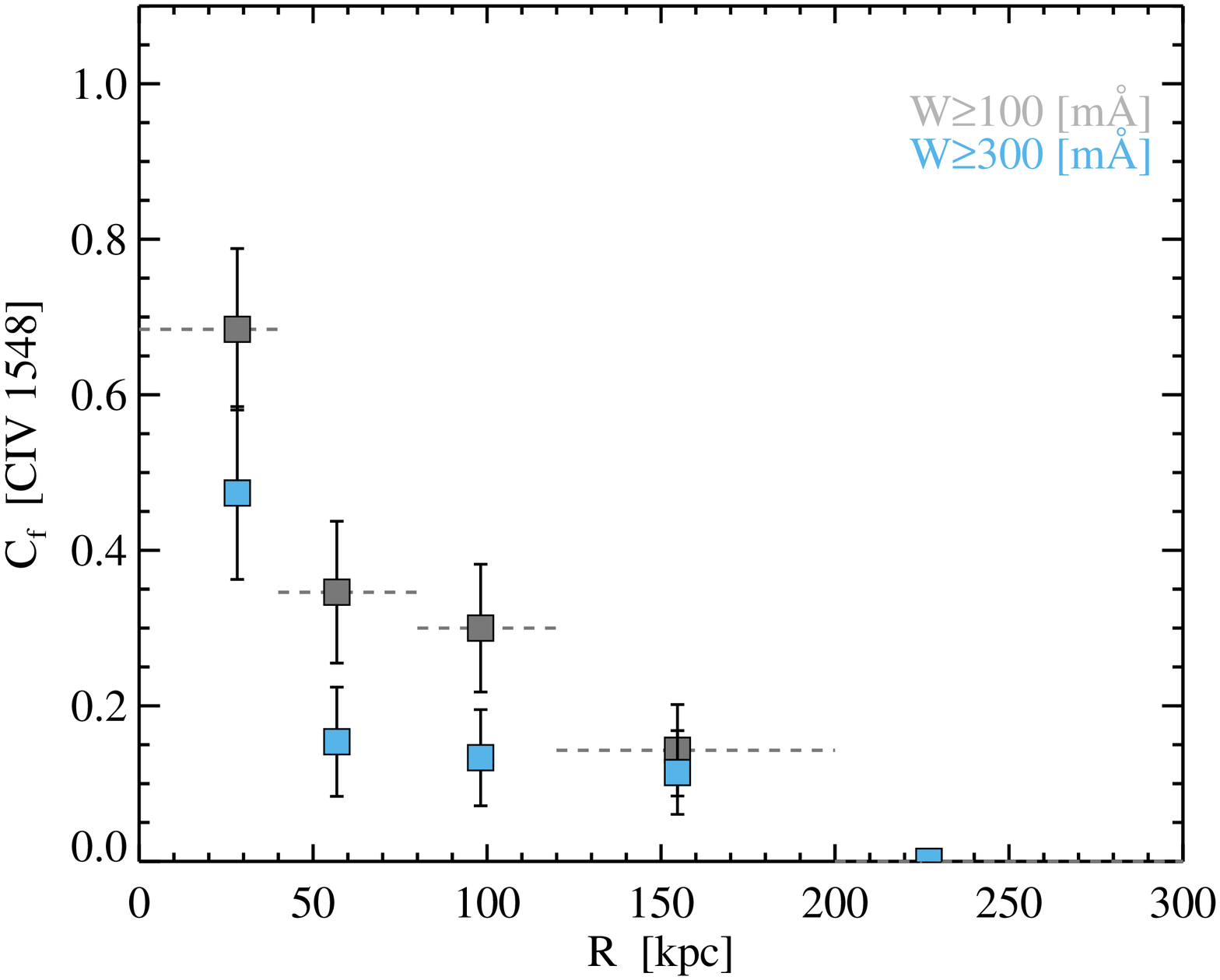}   
\caption{\CIV absorption profile compiled from available literature data These galaxies are not constrained in terms of their masses, star-formation rates, and redshifts. Top Four Panels:- 1-D \CIV absorption profile as a function of R from the literature. The blue and red points are star-forming and passive galaxies from the COS-Dwarfs (this work) and COS-Halos \citep{Tumlinson2013} survey. The gray points are from \cite{chen2001} at z$\sim$0.4 and green points are from \cite{Borthakur2013} at z$\leq 0.2$. The purple points are data from stacked spectra analysis of \citep{Steidel2010} at z$\approx$ 2.2 and orange points are data from  \citep{Prochaska2013b} mapping the CGM around quasar host galaxies at z$\approx$ 2. All the open arrow symbols are 2$\sigma$ non detections.  Bottom Panel:- The 1-D covering fraction profile as a function of R of all the $z <0.5$, literature data with W$ \geq $ 100{m\AA} (gray squares) and with  W$ \geq $ 300{m\AA} (blue squares). The error bars represent the 68\% confidence intervals. Horizontal dashed lines are range bars indicating the width of the radial bins.}
\label{fig: civ_literature}
\end{figure*}

In the top panels of Figure \ref{fig: civ_literature}, we present the \CIV absorption observed in the COS-Halos survey (two detections and one non-detection) and this work. The blue and red data points correspond to star-forming and passive galaxies, respectively. All the open symbols with arrows indicate 2$\sigma$ non-detections. 

Using HST Key Project data, \cite{chen2001} characterized  the \CIV absorption profile around 50 $z$ $\approx$ 0.4 galaxies with background quasar lines of sight passing within 300 kpc of the host galaxies. In Figure \ref{fig: civ_literature}, these measurements are represented as gray points in the top left and the middle left panels. They found \CIV absorption out to R $ \sim$ 100 kpc with abrupt boundaries between \CIV absorbing and non-absorbing regions. They also reported that the \CIV absorption strength does not depend strongly on galaxy surface brightness, redshift, or morphological types. They found in their sample 28\% (14 out of 50) of the galaxies are associated with \CIV absorption. 

In another study, \cite{Borthakur2013}, targeted a sample of 20 $z$ $<$ 0.2 galaxies and constrained the \CIV absorption in 17 of these galaxies within R $ \leq$ 200 kpc. They found that 4 out of 5 star-bursting galaxies exhibit strong \CIV absorption out to 200 kpc. These are represented as green points in the top left and middle right panels of Figure \ref{fig: civ_literature}. 

At higher redshifts, \cite{Steidel2010} used stacked background galaxy spectra to characterize the \CIV absorption around 512 foreground z $\approx$ 2.2 galaxies out to 125 kpc. They found that the \CIV absorption strength falls off sharply between 50-100 kpc. In Figure \ref{fig: civ_literature}, the purple data points in top left and middle right panels represent these measurements. In the \cite{Steidel2010} study, the spectral resolution was too low to resolve the \CIV doublet. Hence the quoted  \CIV absorption strength was summed over both lines of the doublet. Here we use half of the quoted value in \cite{Steidel2010} to compare with the \civ absorption strength. \cite{Prochaska2013b} use a subset of 428 quasars that show \lya absorption to probe the CGM around z $\approx$ 2 quasar host galaxies. They find that \CIV absorption around quasar hosts extends farther than that observed around LBGs at the same redshifts and the covering fraction of strong \CIV absorption extends beyond 200 kpc. We select all the \CIV absorbers within 800 \kms of the systemic redshift of the quasar host galaxy and plot them in Figure \ref{fig: civ_literature} (orange points).

The top left panel of Figure \ref{fig: civ_literature} combines all of the above together. It should be noted that these galaxies represent a very heterogeneous sample of galaxies, spanning many decades in mass, varying star formation rates, and redshifts. Excluding the star-bursting systems of \cite{Borthakur2013} and the \cite{Prochaska2013b} study, the average \CIV absorption profile remains quite unchanged from $z\; \approx $ 2.2 \citep{Steidel2010}, to z $\approx$ 0.4 \citep{chen2001}, all the way to z $\sim$ 0 (COS-Dwarfs). We can detect \CIV absorption out to 100 kpc for all these galaxies.

In the bottom panel of Figure \ref{fig: civ_literature}, we present the \CIV absorption covering fraction for the $z < 0.5$ studies of COS-Halos, COS-Dwarfs, \cite{Borthakur2013} and \cite{chen2001}, as a function of impact parameter. The horizontal dashed lines are range bars indicating the width of the bins. We find that the \CIV absorption covering fraction drops from 70\% (W $\geq$ 100 m{\AA} ) at R $ \leq $ 50 kpc to 30\% at  R $ \approx $ 100 kpc. The 15\% covering fraction seen at extended radius of R $ >$ 150 kpc is primarily contributed by the star-bursting systems of \cite{Borthakur2013}. However, there are observations of individual \CIV absorption line systems, where the host galaxy is observed to be at high impact parameters \citep{Tripp2006,Burchett2013}  and may indicate diverse origins of \CIV absorbers from that described here. This combined dataset indeed suggests that for regular main-sequence non-starburst galaxies, \CIV absorption can been detected out to $\approx$ 100 kpc from z $\sim$ 2 to z $\sim$0 at these detection limits.

\section{Conclusions}
In this work, we have mapped the spatial distribution of \CIV gas around a set of 43 galaxies having mass $\rm{M_{*} \leq 10^{10}}$ \msun at $z <$ 0.1. The main results of this study are as follows.

\begin{itemize}
 \item \CIV absorption is detected out to $\approx$ 100 kpc from the galaxies.  \CIV absorption strength drops off with projected galactocentric radius from the associated galaxy; beyond 0.5$\rm{R_{vir}}$ no \CIV absorption is detected at our sensitivity limits ($\sim$ 50 - 100 m{\AA}).

 \item The \CIV absorption is patchy even at small impact parameters. At $\rm{R \; \leq}$ 60 kpc, 9 out of 17  (60\%) galaxies are associated with \CIV absorption;  at $\rm{R \; \leq}$ 150 kpc 17 out of 40 (43\%) galaxies are associated with \CIV absorption. The patchiness is also evident in the scatter of \CIV absorption strength for the detected absorbers (standard deviation = 208 m{\AA}).

  \item For star-forming galaxies, the covering fraction of  $W \geq $100 {m\AA} \civ absorption is nearly unity at R/\rvir $\rm{\leq}$ 0.2, falling off  to 60\% at {0.4 R/\rvir} and then to zero at  R/\rvir $\gtrsim $ 0.5. 
   
 \item   We find that strong \CIV absorbers are primarily detected around star-forming galaxies and report a correlation between the detected \CIV absorption strength and sSFR of the host galaxies within 0.5\rvir. We find that within {0.5\rvir}, the \CIV detection probability around star-forming galaxies is P = 0.63 $\pm$0.096 and around passive galaxies is P = 0.25 $\pm$0.14. We reject the null hypothesis that there is no correlation between \CIV absorption strength and host galaxy sSFR at the 95\% confidence level.
  
 \item  The detected \CIV absorbers are kinematically consistent with being bound to the dark matter halos of their host galaxies. The absorption centroids cluster around the systemic zero velocity of their host galaxies with a median velocity of 13 {\kms} and a standard deviation of 50 {\kms}.
 
 \item We estimate the minimum carbon mass in the CGM of these galaxies, and find that there is at least $\gtrsim 1.2 \times 10^{6}$ \msun of carbon within 110 kpc  of these galaxies. This is comparable to the total carbon mass in the ISM of these galaxies, and more than the total carbon mass contained in the stars of these galaxies.  Amongst star-forming galaxies, the CGM around low mass galaxies ($\rm{8 \leq \log M_*/M_{\odot} \leq 9.5}$) can have a carbon mass of at least $\gtrsim 0.5 \times 10^{6}$ \msun, and around $\rm{9.5 \leq \log M_*/M_{\odot} \leq 10}$ galaxies the carbon mass is at least  $\gtrsim 2.6 \times 10^{6}$ \msun. This is a conservative lower limit, and the CGM carbon mass could easily be factor of six higher. 
 
  \item Combining the cosmic number density of the galaxies in our sample from the galaxy stellar mass function, with the observed \CIV absorption cross-section could account for $\approx$ 60\% of all intervening \CIV absorption with (W > 100 {m\AA} and W > 300 {m\AA}) and $\approx$ 64\% (W > 400 {m\AA}) \CIV absorbers in the low-$z$ Universe, observed along QSO lines of sight.

\item We compare the observed and simulated galaxies at the same stellar mass in three flavors of feedback models in hydrodynamical simulations. Comparing the hydrodynamical simulations with and without winds, we conclude that the observed metallic content of the CGM around \subl galaxies cannot be explained by tidal debris and ram pressure stripping alone, and strong outflows are required to explain the observations. We find that the energy-driven wind model (ezw) is a better overall fit to the present star formation rates and CGM \CIV covering fraction measurements as compared to the constant velocity wind model (cw). 
 
\end{itemize}
In summary, we examined the CGM of \subl galaxies traced by \CIV absorption. The detection of CGM carbon masses (traced by C IV) comparable to the ISM carbon mass of these low mass galaxies suggest that a substantial fraction of the baryon budget might be hidden in the CGM of these galaxies. In future work, we shall focus on the other ionization species to characterize the CGM of these \subl galaxies and extend the baryon census began by the COS-Halos survey to these low masses.

\section{Acknowledgement}
Support for program GO12248 was provided by NASA through a grant from the Space Telescope Science Institute, which is operated by the Association of Universities for Research in Astronomy, Inc., under NASA contract NAS 5-26555. BDO was supported by HST grant HST-AR-12841. Support for the simulations presented in this work was provided by the Ahmanson Foundation.

\input{line_table.tex}

\begin{deluxetable*}{ccccc}
\tablewidth{0pt}
\tablecaption{Minimum Carbon Mass Estimates: }
\tablehead{
\colhead{ Title  }&
\colhead{$\log M_{*}$/\msun }&
\colhead{ $\rm{M_{carbon}}$/\msun \tablenotemark{a}}&
\colhead{ $\rm{M_{carbon}}$/\msun\tablenotemark{b}}&
}
\startdata

   &     8 - 9.5    &      0.4 $\times \;10^{6}$  &      0.4 $\times \;10^{6}$ \\
All galaxies &  9.5 -10  & 0.9 $\times \;10^{6}$  &      1.9 $\times \;10^{6}$ \\  
\\
\hline
\\
   &     8 - 9.5    &      0.5 $\times \;10^{6}$  &      0.5 $\times \;10^{6}$ \\
Star-forming galaxies &  9.5 -10  & 1.1 $\times \;10^{6}$  &      2.6 $\times \;10^{6}$ \\  
\enddata
\label{mass_table}
\tablenotetext{a}{  AOD $N_{CIV}$}
\tablenotetext{b}{ Voigt Profile fitted $N_{CIV}$}
\end{deluxetable*}

\renewcommand\bibsection{}
\bibliographystyle{thesis_bibtex}

\bibliography{mybibliography}

\appendix
HST-COS spectra of the detected 17 \CIV absorbers over-plotted with their corresponding Voigt profile fits are shown below. 

\begin{figure*}
\centering
\subfloat[]{\label{fig:1}\includegraphics[width=.8\textwidth]{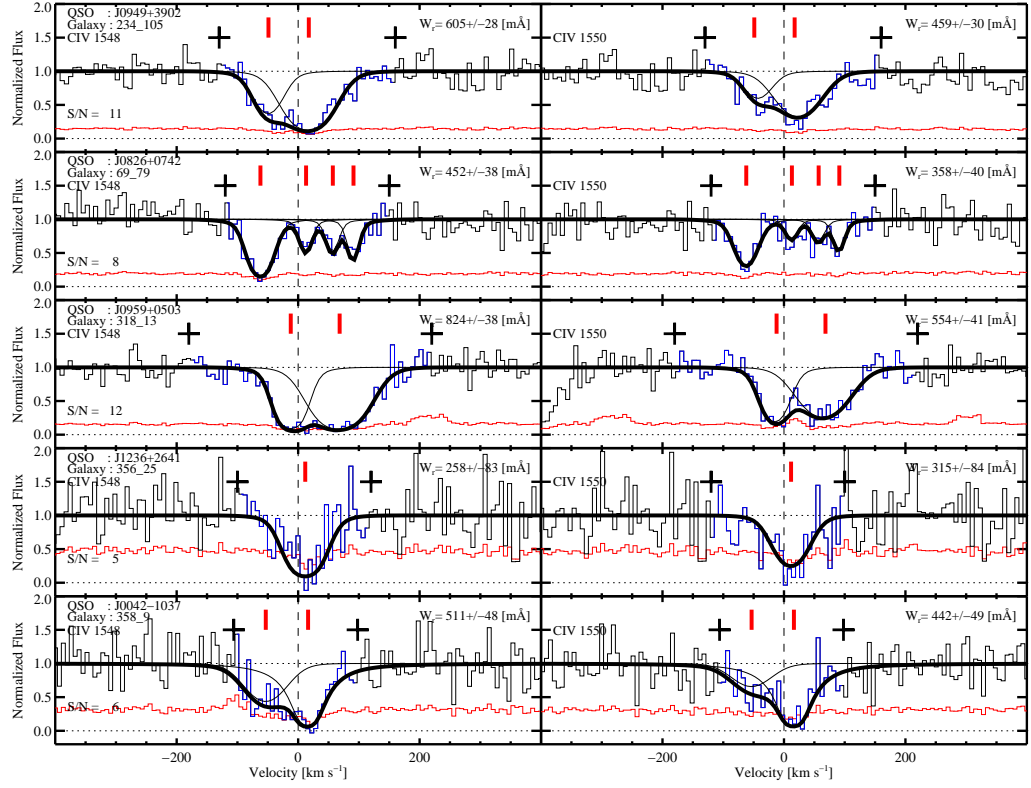}}
\qquad
\subfloat[]{\label{fig:1}\includegraphics[width=.8\textwidth]{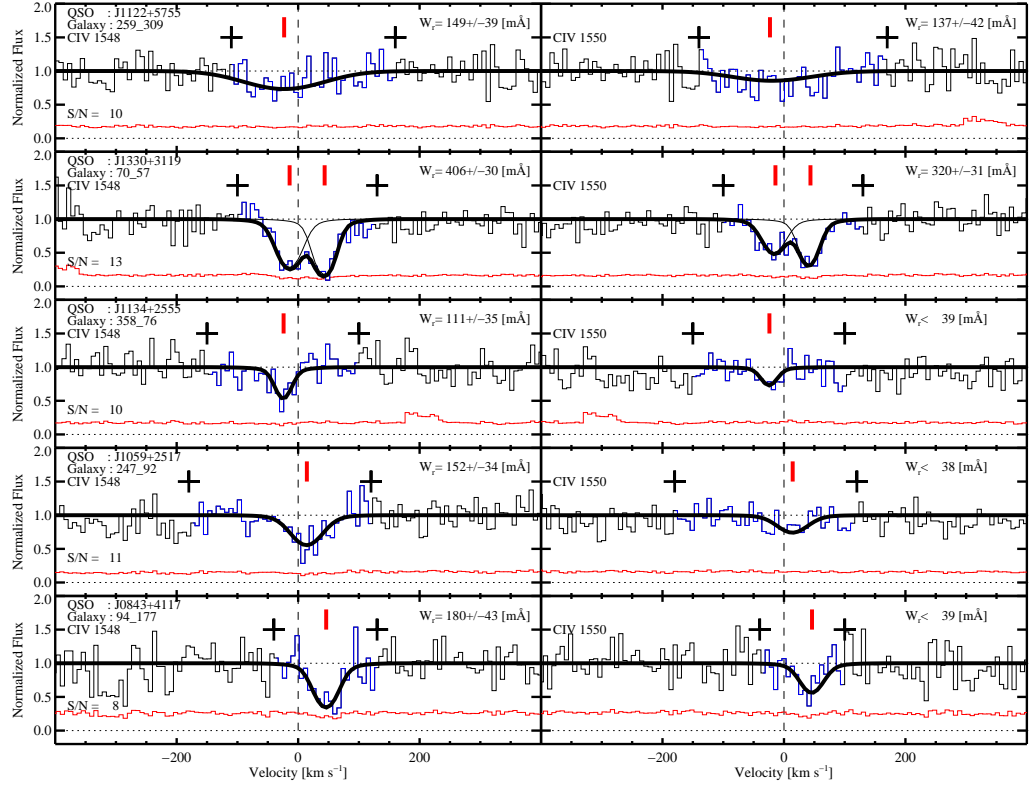}}
\qquad
\caption{HST-COS quasar spectra of the \CIV absorption doublets with their corresponding Voigt profile fits (solid black line). The vertical red ticks indicate the centroids of individual Voigt profile components and the black crosses show the velocity range over which the profile was integrated to compute their equivalent widths. For each system their rest frame equivalent widths and the S/N of the spectrum at \CIV rest frame are shown. }
\label{fig:cont}
\end{figure*}

\begin{figure*}
\ContinuedFloat
\centering
\subfloat[]{\label{fig:1}\includegraphics[width=0.8\textwidth]{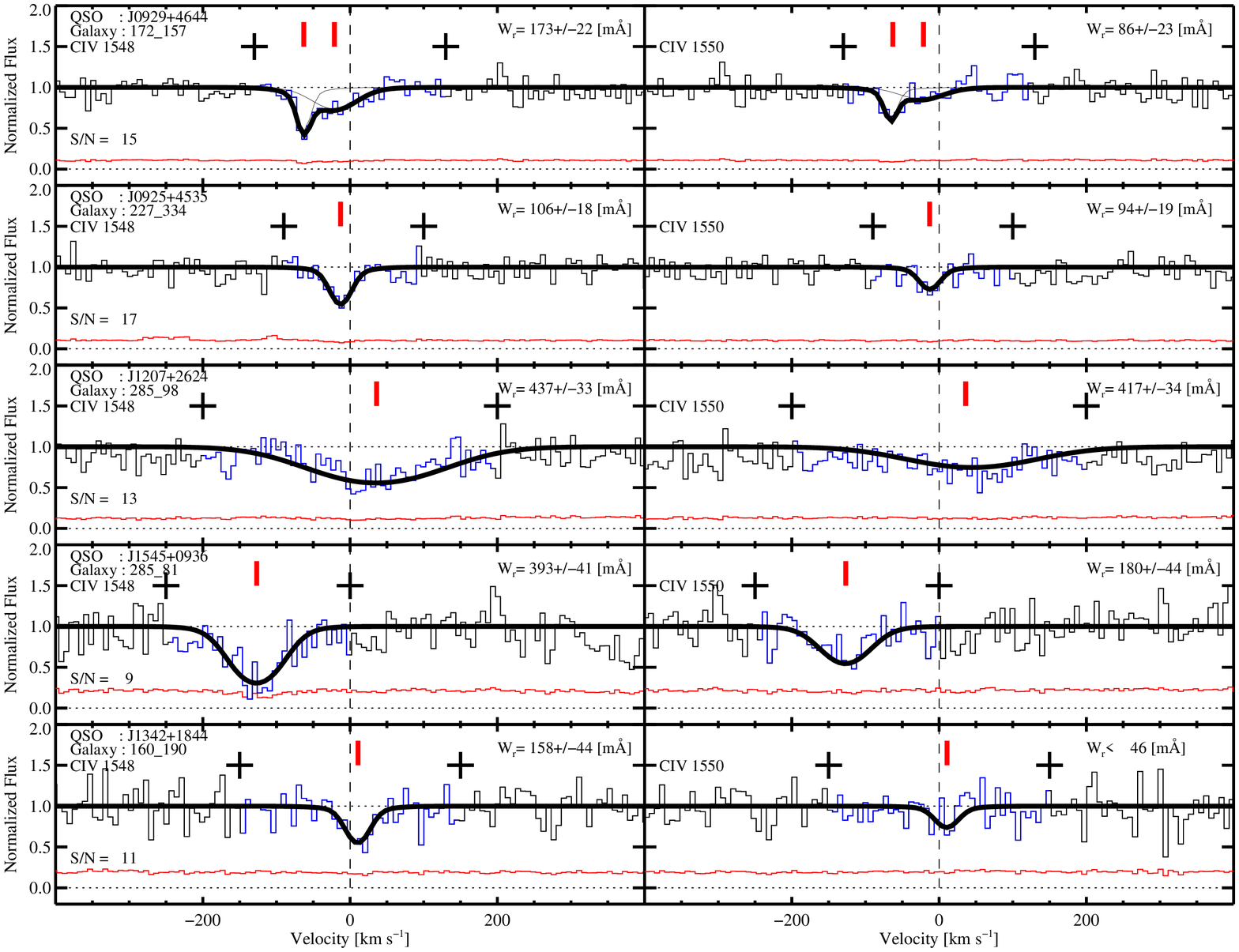}}
\qquad
\subfloat[]{\label{fig:1}\includegraphics[bb=0 0 778 284,width=0.8\textwidth]{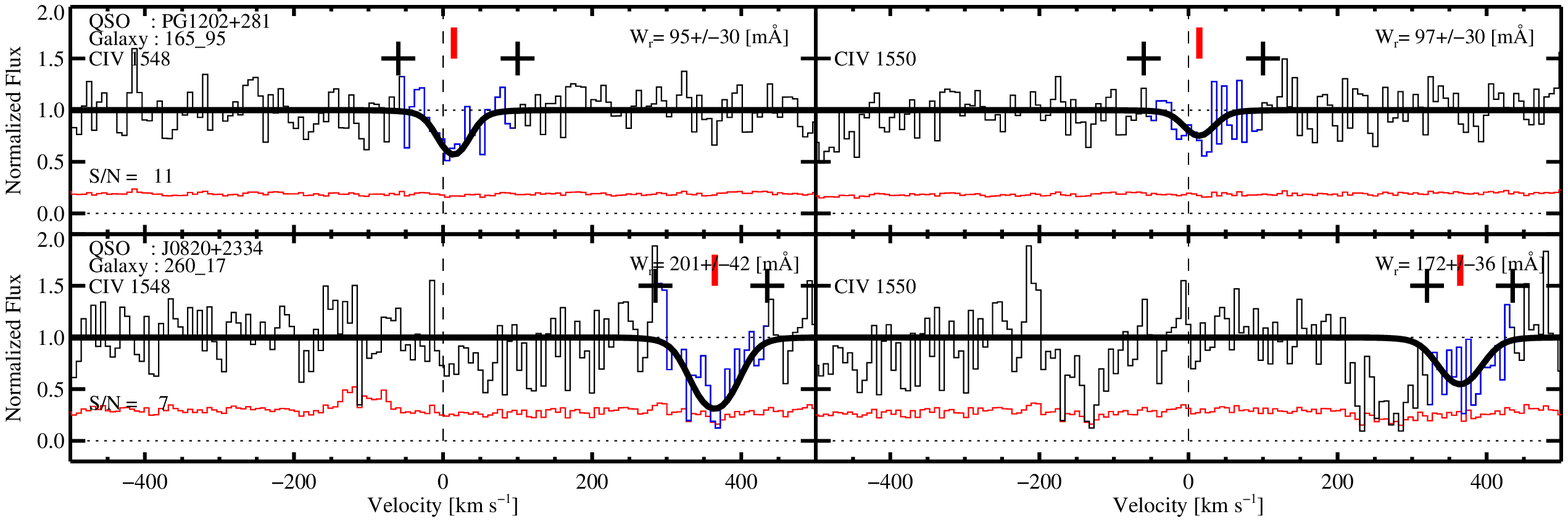}}

\caption{continued}
\label{fig:cont}
\end{figure*}

\end{document}

%% file: line_table.tex
\begin{deluxetable*}{cccccccccccccc}
\tablecolumns{14}
\tablewidth{0pt}
\tablecaption{COS-Dwarfs Galaxy-CIV Measurements: }
\tablehead{
\colhead{QSO Name}&
\colhead{Galaxy\tablenotemark{a}}&
\colhead{Galaxy}&
\colhead{Galaxy}&
\colhead{$z_{\rm{sys}}$}&
\colhead{L/L*\tablenotemark{b}}&
\colhead{$\log M_{*}$}&
\colhead{ R \tablenotemark{c}}&
\colhead{$\rm{R_{vir}}$\tablenotemark{d}}&
\colhead{log sSFR}&
\colhead{$\rm{\log (N_{CIV}}$)\tablenotemark{e}}&
\colhead{$\rm{W_r}$\tablenotemark{f}}&
\colhead{$\rm{\phi}$ \tablenotemark{g}}&
\colhead{\CIV 3$\sigma$ }\\
\colhead{}&
\colhead{}&
\colhead{$\alpha$  [J2000]}&
\colhead{$\delta$  [J2000]}&
\colhead{}&
\colhead{}&
\colhead{[$M_{\odot}$]}&
\colhead{[kpc]}&
\colhead{[kpc]}&
\colhead{[$\rm{yr^{-1}}$]}&
\colhead{[$\rm{cm^{-2}}$]}&
\colhead{[m\AA]}&
\colhead{[Deg]}&
\colhead{detection limit [m\AA]}
}
\startdata
J0929+4644 & 172\_157  &  09:29:11.71 &   +46:41:47.2 &   0.017 &   0.019 &     8.5 &    52 &   132 &    -8.7 &   13.73 $\pm$ 0.05     &   173 $\pm$ 22 & 50 &   30  \\
J0925+4535 & 227\_334  &  09:25:30.98 &   +45:31:57.8 &   0.014 &   0.111 &    10.0 &    95 &   259 &   -10.3 &   13.56 $\pm$ 0.06     &   106 $\pm$ 18 &  9 &   30  \\
J1207+2624 &  285\_98  &  12:07:13.89 &   +26:24:55.4 &   0.048 &   0.066 &     9.7 &    90 &   225 &   -10.0 &   14.19 $\pm$ 0.03     &   437 $\pm$ 33 & 11 &   36  \\
J1545+0936 &  285\_81  &  15:45:48.20 &   +09:36:42.2 &   0.055 &   0.181 &     9.8 &    84 &   232 &    -9.8 &   14.14 $\pm$ 0.05     &   393 $\pm$ 41 & 55 &   57  \\
J1342+1844 & 160\_190  &  13:42:51.43 &   +18:41:44.5 &   0.027 &   0.052 &     9.4 &   101 &   202 &   -10.0 &   13.71 $\pm$ 0.11     &   158 $\pm$ 44 & 65 &   54  \\
J1122+5755 & 259\_309  &  11:22:06.67 &   +57:54:45.3 &   0.011 &   0.029 &     9.1 &    65 &   182 &   -10.1 &   13.74 $\pm$ 0.08     &   149 $\pm$ 39 & 86 &   51  \\
J1330+3119 &   70\_57  &  13:30:57.52 &   +31:19:50.2 &   0.034 &   0.045 &     9.5 &    37 &   206 &   -10.1 &   14.27 $\pm$ 0.03     &   406 $\pm$ 30 &  7 &   51  \\
J1134+2555 &  358\_76  &  11:34:57.46 &   +25:56:44.6 &   0.032 &   0.080 &     9.7 &    47 &   224 &    -9.9 &   13.58 $\pm$ 0.12     &   111 $\pm$ 35 & 78 &   48  \\
J1059+2517 &  247\_92  &  10:59:52.49 &   +25:16:33.5 &   0.021 &   0.117 &     9.9 &    37 &   252 &   -10.1 &   13.74 $\pm$ 0.08     &   152 $\pm$ 34 & 85 &   42  \\
J0843+4117 &  94\_177  &  08:44:05.20 &   +41:17:26.4 &   0.030 &   0.040 &     9.6 &   103 &   216 &    -9.9 &   13.84 $\pm$ 0.10     &   180 $\pm$ 43 & 22 &   75  \\
J0949+3902 & 234\_105  &  09:49:45.57 &   +39:01:01.9 &   0.018 &   0.064 &     9.7 &    37 &   233 &    -9.7 & $>$  14.54             &   605 $\pm$ 28 &  9 &   42  \\
J0826+0742 &   69\_79  &  08:26:38.51 &   +07:43:16.4 &   0.052 &   0.085 &     9.7 &    77 &   222 &    -9.6 & $>$  14.39             &   452 $\pm$ 38 & 63 &   54  \\
J0959+0503 &  318\_13  &  09:59:15.04 &   +05:04:05.0 &   0.059 &   0.088 &    10.0 &    14 &   247 &    -9.9 & $>$  14.69             &   824 $\pm$ 38 & 56 &   42  \\
J1236+2641 &  356\_25  &  12:36:03.91 &   +26:42:01.4 &   0.062 &   0.043 &     9.4 &    28 &   193 &    -9.7 & $>$  14.38             &   258 $\pm$ 83 & 50 &  144  \\
J0042-1037 &   358\_9  &  00:42:22.27 &   -10:37:35.2 &   0.095 &   0.039 &     9.6 &    15 &   200 &   -10.3 & $>$  14.55             &   511 $\pm$ 48 & 27 &   87  \\
PG1202+281 &  165\_95  &  12:04:43.87 &   +27:52:39.2 &   0.051 &   0.096 &    10.0 &    92 &   254 & $<$  -12.1 &   13.58 $\pm$ 0.10  &    95 $\pm$ 30 & 59 &   54  \\
J0820+2334 &  260\_17  &  08:20:22.99 &   +23:34:47.4 &   0.095 &   0.046 &     9.8 &    29 &   218 & $<$  -10.9 & $>$  14.10          &   201$\pm$ 42  & 50 &   87  \\
J0809+4619 & 257\_269 &   08:08:42.75 &   +46:18:29.0 &   0.024 &   0.030 &     9.0 &   125 &   169 &    -9.6 & $<$  12.88             & $<$   14       & 52 &   30  \\
J1327+4435 & 122\_131 &   13:27:14.52 &   +44:33:54.3 &   0.048 &   0.089 &     9.8 &   119 &   229 &    -9.5 & $<$  13.25             & $<$   19       & 32 &   45  \\
J0912+2957 &  20\_223 &   09:12:41.57 &   +30:00:53.9 &   0.023 &   0.062 &     9.8 &   102 &   240 &   -10.0 & $<$  12.97             & $<$   18       & 42 &   39  \\
J0947+1005 & 135\_580 &   09:48:00.79 &   +09:58:15.4 &   0.010 &   0.014 &     9.0 &   120 &   172 &   -10.0 & $<$  13.02             & $<$   20       & 73 &   45  \\
PG1049-005 &  316\_78 &   10:51:47.77 &   -00:50:20.3 &   0.039 &   0.110 &     9.6 &    58 &   219 &    -9.9 & $<$  12.98             & $<$   16       & 53 &   36  \\
J1521+0337 & 252\_124 &   15:21:31.73 &   +03:36:51.8 &   0.036 &   0.087 &     9.5 &    87 &   206 &    -9.5 & $<$  13.56             & $<$   58       & 69 &  120  \\
J1451+2709 & 184\_526 &   14:51:05.82 &   +27:00:42.3 &   0.013 &   0.008 &     8.3 &   135 &   114 &    -9.8 & $<$  13.09             & $<$   18       & 81 &   39  \\
J1342+0505 & 210\_241 &   13:41:58.41 &   +05:01:55.7 &   0.025 &   0.038 &     9.4 &   116 &   201 &    -9.9 & $<$  13.35             & $<$   17       & 73 &   39  \\
J1211+3657 & 312\_196 &   12:11:02.54 &   +36:59:53.9 &   0.023 &   0.147 &    10.1 &    90 &   272 &    -9.8 & $<$  13.17             & $<$   29       & 89 &   45  \\
J1121+0325 &  73\_198 &   11:21:26.95 &   +03:26:41.6 &   0.023 &   0.174 &    10.1 &    89 &   277 &   -10.2 & $<$  13.45             & $<$   44       &  7 &  114  \\
J1001+5944 &  87\_608 &   10:02:23.03 &   +59:44:34.5 &   0.011 &   0.018 &     8.7 &   135 &   148 &    -9.7 & $<$  12.89             & $<$   15       & 76 &   30  \\
J0155-0857 & 329\_403 &   01:55:16.26 &   -08:51:15.4 &   0.013 &   0.015 &     9.0 &   105 &   169 &    -9.8 & $<$  13.15             & $<$   28       & 24 &   51  \\
J0310-0049 & 124\_197 &   03:10:38.64 &   -00:51:43.6 &   0.026 &   0.008 &     8.5 &   101 &   129 &    -9.9 & $<$  13.04             & $<$   21       & 70 &   48  \\
J0242-0759 &  84\_223 &   02:43:05.80 &   -07:58:53.3 &   0.029 &   0.040 &     8.9 &   126 &   166 &    -9.6 & $<$  13.23             & $<$   34       & 32 &   72  \\
J1059+1441 & 316\_200 &   10:59:35.63 &   +14:44:07.7 &   0.010 &   0.006 &     8.2 &    41 &   102 &    -9.1 & $<$  13.06             & $<$   16       & 19 &   36  \\
J1357+1704 &  93\_248 &   13:57:29.95 &   +17:04:29.2 &   0.026 &   0.042 &     9.3 &   124 &   196 &    -9.7 & $<$  12.93             & $<$   17       & 19 &   39  \\
J0946+4711 & 198\_218 &   09:46:14.61 &   +47:08:03.3 &   0.015 &   0.080 &     9.6 &    66 &   222 &    -9.9 & $<$  13.71             & $<$   59       & 18 &  156  \\
J1022+0132 &  337\_29 &   10:22:18.22 &   +01:32:45.4 &   0.074 &   0.024 &     9.1 &    39 &   167 &   -10.1 & $<$  13.31             & $<$   45       & 71 &   72  \\
J1616+4154 &  327\_30 &   16:16:47.99 &   +41:54:41.3 &   0.104 &   0.041 &     9.2 &    55 &   172 &    -9.4 & $<$  13.47             & $<$   39       & 84 &   63  \\
J1356+2515 & 182\_159  &  13:56:25.02 &   +25:12:44.2 &   0.032 &   0.038 &     9.6 &    97 &   212 & $<$  -12.1 & $<$  13.27          & $<$   35       & 87 &   72  \\
J1210+3157 &  65\_308  &  12:10:59.71 &   +31:59:12.2 &   0.022 &   0.022 &     9.3 &   134 &   197 & $<$  -11.7 & $<$  13.11          & $<$   25       & 49 &   57  \\
J1117+2634 & 114\_210  &  11:18:08.60 &   +26:32:50.1 &   0.028 &   0.063 &     9.8 &   114 &   239 & $<$  -12.1 & $<$  13.28          & $<$   32       & 16 &   72  \\
J1104+3141 &  211\_65  &  11:04:04.25 &   +31:40:15.1 &   0.047 &   0.079 &     9.7 &    58 &   225 & $<$  -10.9 & $<$  13.23          & $<$   29       & 16 &   39  \\
J1342+3829 & 322\_238  &  13:42:18.70 &   +38:32:12.2 &   0.012 &   0.031 &     9.2 &    54 &   190 & $<$  -11.0 & $<$  13.07          & $<$   19       &  6 &   42  \\
J0212-0737 & 334\_153  &  02:12:13.91 &   -07:35:01.1 &   0.018 &   0.022 &     9.0 &    53 &   170 & $<$  -10.7 & $<$  12.95          & $<$   19       & 21 &   51  \\
J1103+4141 &   13\_58  &  11:03:14.15 &   +41:42:51.7 &   0.030 &   0.029 &     9.5 &    33 &   206 & $<$  -11.5 & $<$  13.34          & $<$   25       & 77 &   45  \\
   \enddata
   \vspace{-0.2cm}
\label{line_table}
\tablenotetext{a}{We label the galaxies by the position angle with respect to the QSO, N through E, and with the angular separation in arcsec.}
\tablenotetext{b}{Galaxy luminosity in terms of L*, where L* is given by an R-band absolute magnitude of -21.12 \citep{Blanton2001}. }
\tablenotetext{c}{Impact parameter in kpc.}
\tablenotetext{d}{Virial radius in kpc.}
\tablenotetext{e}{Limits on $N_{CIV}$ are 2 $\sigma$, AOD column Densities}
\tablenotetext{f}{\civ rest frame equivalent widths. Limits on $W_r$ are 1$\sigma$.}
\tablenotetext{g}{Azimuthal angle with respect to the galaxy's projected minor axis.}
\end{deluxetable*}